\documentclass[preprint, 1p]{elsarticle}

\usepackage{simpler-wick}

\usepackage{ulem}
\usepackage[T1]{fontenc}
\usepackage[utf8]{inputenc}
\usepackage[german,english]{babel}
\usepackage{epsfig}
\usepackage{amssymb}
\usepackage{amsmath}
\usepackage{braket}
\usepackage{bm}
\usepackage{color}
\usepackage{times}
\usepackage[colorlinks,bookmarks=false,citecolor=blue,linkcolor=red,urlcolor=blue]{hyperref}

\definecolor{darkGreen}{rgb}{0,0.4,0}

\usepackage{soul}

\DeclareMathOperator{\tr}{Tr}

\newcommand{\beq}{\begin{equation}}
\newcommand{\eeq}{\end{equation}}
\newcommand{\bea}{\begin{eqnarray}}
\newcommand{\eea}{\end{eqnarray}}

\newcommand{\blue}{}

\begin{document}
\setstcolor{red}
%%%%%%%%%%%%%%%%%%%%%%%%%%%%%%%%%%%%%%%%%%%%%%%%%%%%%%%%%%%%%%%%%%%%%%%%%%%%%%
\title{Exploring many-body localization \\ in  quantum systems coupled to an environment\\ via Wegner-Wilson flows 
}
%%%%%%%%%%%%%%%%%%%%%%%%%%%%%%%%%%%%%%%%%%%%%%%%%%%%%%%%%%%%%%%%%%%%%%%%%%%%%%

\author{Shane P. Kelly} %\footnote{jamirmarino@fas.harvard.edu}
\address{Theoretical Division, Los Alamos National Laboratory, Los Alamos, New Mexico 87545, USA \\
Physics and Astronomy Department, University of California Riverside, Riverside, California 92521, USA}
\author{Rahul Nandkishore} %\footnote{jamirmarino@fas.harvard.edu}
\address{Department  of  Physics  and  Center  for  Theory  of  Quantum  Matter,
University  of  Colorado,  Boulder,  CO  80309, USA\\
Kavli Institute for Theoretical Physics, University of California, Santa Barbara, CA 93106-4030, USA}
\author{Jamir Marino} %\footnote{jamirmarino@fas.harvard.edu}
\ead{jamirmarino@fas.harvard.edu}
%\email{jamirmarino@fas.harvard.edu}
\address{Department of Physics, Harvard University, Cambridge MA 02138, USA\\
Department of Quantum Matter Physics, University of Geneva, 1211, Geneve, Switzerland \\
Kavli Institute for Theoretical Physics, University of California, Santa Barbara, CA 93106-4030, USA}

%\noindent

%\begin{abstract}

\begin{abstract} 
    Inspired by recent experiments on many-body localized systems coupled to an environment, we apply a Flow Equation method to study the problem of a disorder chain of spinless fermions, coupled via density-density interactions to a second clean chain of spinless fermions.  
    In particular, we focus on the conditions for the onset of a  many-body localized phase in the clean sector of our model  by proximity to the dirty one.
    We find that a many-body localization proximity effect in the clean component is  established when the density of dirty fermions exceeds a threshold value.
    From the flow equation method we find that, similar to many-body localization in a single chain, the many-body localization proximity effect is also described by an extensive set of local integrals of motion. 
    Furthermore, by tuning the geometry  of the inter-chain couplings, we show that the dynamics of the model is ruled, on intermediate time scales, by an emergent set of quasi-conserved charges. % Hamiltonian with a novel  set of emergent integrals of motion.
 
    %We apply a Flow Equation method to study the \spk{problem of   a many-body localized system} of one-dimensional  spinless fermions, coupled   via  density-density interactions  to a second clean chain of  fermions.  
    %In particular, we focus on the conditions for the onset of a  many-body localized phase in the clean sector of our model  {by proximity} to the dirty one. We find that  a many-body localization proximity effect in the clean component is  established when the density of dirty fermions exceeds a   threshold value, \spk{in a way reminiscent of  recent experiments on many-body localized systems coupled to a bath. }
    %By tuning the control parameters of the model, we establish thresholds for the induction of a many-body localized phase in the clean sector, \spk{using a joint set of emergent integrals of motion for the clean and dirty components as ansatz for the solution of the flow equations.}
    %Furthermore, by tuning the geometry  of the inter-chain couplings, we show that the dynamics of the model can be described, on intermediate time scales, by a Hamiltonian with a novel  set of emergent integrals of motion.
 \end{abstract}
%\end{abstract}

%\begin{multicols}{2}

\date{\today}
\maketitle
\section{Introduction}

The advent of cold gas experiments~\cite{RevModPhys.80.885} has revitalized  interest in  fundamental questions of quantum thermodynamics in isolated  many-body systems.
One of the most intriguing avenues of research is the quest for non-ergodic phases of quantum matter. Examples range from integrable models~\cite{Essler2016, Calabrese16} to quantum scars~\cite{scars} {and include} the prominent example of ergodicity breaking by strong disorder: many-body localization (MBL) ~\cite{Nandkishore-2015, review} 

MBL has been the subject of intense research activity in the last ten years; seminal works have  studied the problem  both in a perturbation treatment~\cite{AGKL, BAA, Mirlinrecent} and with numerical methods~\cite{Pal, Znidaric, OganesyanHuse}, establishing that a localized phase, which exhibits absence of diffusion on long time scales, can survive the presence of many body interactions.
Interest in many-body localisation results from its rich phenomenology: unusual dynamical responses~\cite{nonlocal, mblconductivity},  a novel pattern of quantum entanglement~\cite{Bardarson, Geraedts2016, KhemaniPRX, Chamon, GRN}, the possibility to host  new types of order without equilibrium  counterpart~\cite{LPQO, Pekkeretal2014, VoskAltman2014, hysRevLett.110.067204}, and  connections to the notion of quantum integrability \cite{Bardarson, Serbyn, HNO, Scardicchio, lstarbits, GBN, NonFermiGlasses}.
MBL systems possess an extensive set of quasi-local integrals of motion,  conserved by the unitary dynamics, and preventing full thermalization. Such local degrees of freedom (called localized bits or l-bits) can be constructed via a sequence of local unitary transformations starting from a free Anderson insulator, and represents a form of quantum integrability robust to perturbations.
This property is at the basis of a mathematical proof of the existence of the MBL phase for one-dimensional spin lattice systems    with short-range interactions~\cite{Imbrie}.
 
  MBL is nowadays investigated in experiments with cold gases~\cite{Schreiber2015, Bordia, Choi2016, risp} and superconducting qubits~\cite{exp}.
 The advent of MBL in experimental platforms poses naturally the question of its robustness to the coupling with an external environment~\cite{NGH, NGADP}. 
 A bath is expected to provide sufficient energy and phase-space to facilitate the hopping in an otherwise localized system~\cite{gn, BanerjeeAltman, Altmandeph, Levi, Medv16, avalanches, deroeck17, Chandran17, lorenzo}.
 On the other hand, a recent experiment~\cite{Bloch17} suggests that the clean 'environment' needs to reach  a comparatively large density of particles with respect to the  dirty MBL system in order to act as a thermodynamic environment and induce ergodic behavior.
In order to render the problem treatable, the coupling between a quantum many body system and a bath is usually assumed weak. 
The complementary regime, however, presents an even more interesting scenario: when the back-action on the bath is strong,
and the bath and system are of comparable size,
the 'clean' bath could localize by proximity  to the dirty system -- a phenomenon called 'MBL proximity effect'~\cite{proximity, hyatt, PhysRevB.97.054201}.

Previous work has substantiated the existence of the 'MBL' proximity effect' via perturbative treatments~\cite{proximity,PhysRevB.97.054201} and exact numerics on small system sizes~\cite{hyatt}, while to the best of our knowledge there has been no attempt at constructing  integrals-of-motion, or an {\it l-bit} Hamiltonian, for 'MBL proximity' induced phases.
    Therefore, in this work we investigate the possibility of such a construction by use of the  Wegner-Wilson flow equation method~\cite{Kehrein}.
    Similar to renormalization group approaches to the MBL problem\cite{VHO, PVP, PhysRevLett.119.110604,PhysRevLett.121.140601,PhysRevLett.121.206601, vass}, the Wegner-Wilson flow equation method\cite{Kehrein} constructs a set flow equations  implementing  infinitesimal stepwise diagonalization of the many-body Hamiltonian.
When both the clean and dirty components of the system localize, these equations describe a unitary transformation, in both clean and dirty components, to an l-bit Hamiltonian  which is diagonal in an extensive set of local conserved charges.
By focusing on this regime, one can make an ansatz of the l-bit Hamiltonian that only includes a few relevant many-body terms.
Thus, in addition to being able to study regimes of strong system-bath coupling, the flow equation method is also able to access system's sizes beyond those treatable in exact diagonalization, when disorder is sufficiently strong.

{This approach allows us to establish the existence of the MBL proximity effect, and its consistency with a diagonal {\it l-bit} Hamiltonian of local conserved charges, in a wide range of parameters.}
Of particular note, we identify a regime for the MBL proximity effect complementary to the one explored in the experiment of Ref.~\cite{Bloch17}: {above a certain critical density the dirty system acts effectively as a source of disorder} and induces an MBL phase into the clean component. 
We also focus on novel physical regimes occurring when the geometry of the system-bath coupling is modified.
Specifically, we consider the case of a dirty chain of fermions, coupled every $\delta>1$ sites,  to the clean one (see Fig.~\ref{fig:chainDiag}); the dirty chain  acts as a distribution of impurities placed every $\delta$ sites, {cutting the clean chain into a sequence of emergent integrals of motions}. These conserved charges lead to non-ergodic dynamics on intermediate time scales but are destroyed when interactions between conserved charges becomes effective. At these longer time scales, instead, the dynamics cross over from non-ergodic behavior to thermal behavior.

\subsection{Structure of the paper}

We begin in section~\ref{sec:review} with a review of the Wegner-Wilson flow Equation technique for a single chain and discuss how such a technique provides access to local conserved charges and an {\it l-bit} Hamiltonian.
Then, in section~\ref{sec:twoChainIntro}, we generalize the approach for the two-chain problem and detail how to identify the parameter space where the MBL proximity effect is reliably described by an {\it l-bit} Hamiltonian.
In section~\ref{sec:numResults}, we present the numerical solution to the flow equations in the case of two chains of equal length.
Here, we demonstrate the stability of the MBL proximity effect, construct a qualitative phase diagram and present the numerically computed {\it l-bit} couplings.
In section~\ref{sec:flowEQMainTxt}, we describe in greater detail the truncations made by the two-chain {\it l-bit} ansatz and sketch the derivation of the differential equations defining the FE unitary transformation.
In section~\ref{sec:novelGeometry}, we apply the method developed in the first sections to a novel geometry for the system-bath coupling, and discuss a novel relaxation process.
We  conclude with a quick overview on  relevant experiments and possible future directions in section~\ref{sec:discussion}.

\section{Flow Equation Approach For a Single Chain}
\label{sec:review}
The key idea of the FE approach is to introduce a family of unitary transformations, $U(l)$, parameterised by a 'renormalization group' scale, $l$,  and generated by the anti-hermitian operator, $\eta(l)$, via the relation, $U(l)=T_l\exp\left(\int\eta(l)dl\right)$. The fixed point of the FE procedure in the $l\to\infty$ limit, is a diagonal Hamiltonian with dressed couplings. Operators, $O(l)$, flow according to the equation $\frac{d O}{d l}=[\eta(l),O(l)]$. 
A customary procedure for constructing $\eta(l)$ is to first separate the Hamiltonian into its diagonal, $H_0(l)$, and off-diagonal, $V(l)$ parts.
Then, the generator is constructed as $\eta(l)\equiv[H_0(l),V(l)]$ which guarantees vanishing off-diagonal terms at the fixed point, $l\to\infty$~\cite{Wegner1994}.
Typically, the solution of an interacting quantum many-body system via the FE approach would require a broad set of variational parameters keeping track of the nested hierarchies of multi-particles correlations.

However, in the case of MBL systems, a guiding insight in fixing the {variational} ansatz for the flow equations comes from the {l-bit picture}~\cite{Pekker17, Thomson17}, which provides a method to numerically solve the flow in an efficient way: only the first leading terms describing pairwise interactions between the {\it l-bits} are retained, while higher order effects are truncated and discarded.
This represents an excellent description as long as the system is strongly localized. % into the delocalized phase.
Given this ansatz  for $H(l)$, the flow of the couplings is readily given by the solution of $\frac{d H}{d l}=[\eta(l),H(l)]$.
In other words, the flow brings the Hamiltonian of a single disordered fermionic wire (for instance, $H_d$ in Eq.~\eqref{eq:bare}) into an effectively diagonal one at the fixed point 
\begin{equation}\label{eq:trunc1}
    \mathcal{H}(\infty)=\sum_i h_i(\infty) n_i+\sum_{i,j}\Delta_{ij}(\infty)n_in_j.
\end{equation}
%with the dressed coupling at $l\to\infty$, given by $\tilde{h}_i=h_i(\infty)$ and $\tilde{\Delta}_{ij}=\Delta_{ij}(\infty)$.
This, in turn, shows that the FE method effectively brings the Hamiltonian into an {l-bit  basis}, with couplings between the integrals-of-motion that decay in space as $\Delta_{ij}(\infty)_{ij}\propto\exp(-|i-j|/\xi)$. The values of $h_{i}(\infty)$ and $\Delta_{ij}(\infty)$ depend on the specific disorder realization.  Therefore, to consider disorder averaged quantities, the flow equations must be computed independently for each disorder realization.

In addition to extracting  the conserved charges and l-bit Hamiltonian in Eq.~\ref{eq:trunc1}, the FE method can be used to approximate a crossover region from the MBL phase to a delocalized phase~\cite{Thomson17}.
This region is identified with the parameter space where truncation error proliferates.
These errors {indicate} the departure from an MBL phase because they {indicate that the true unitary transformation must contain correlations between local degrees of freedom that are not captured by the ansatz.
Since the growth of correlation between local degrees of freedom is suggestive of delocalization, the proliferation of truncation error is also indicative of a breakdown of the MBL phase.}
In order to measure the truncation error, one calculates the so-called {'second invariant'}~\cite{Thomson17,Monthus_2016}, a quantity conserved by the exact unitary transformation.
Since the truncation breaks the unitarity of the flow, the truncation error is controlled by  changes in second invariant.
%\spk{Furthermore, a large change in the second invariant is suggestive of the breakdown of the MBL phase.}
%\spk{Furthermore, since larger error indicates a local ansatz, associated with the MBL phase, is not a sufficent discription of the diagonal hamiltonian, a large change in the second invariant is suggestive of the breakdown of the MBL phase.}

By setting a small threshold for the change in the second invariant, a {tight bound on the MBL phase region} can be identified with the parameter space where the truncation yields error within  the threshold.
Such analysis performed on the single chain led gives a phase boundary consistent with exact diagonalization~\cite{Thomson17}.
We discuss the second invariant in detail as it pertains to the MBL proximity effect in section~\ref{sec:secndInv}.
%This summarizes the use of the FE procedure for a single chain to construct conserved charges and identify a {tight bound on the MBL phase boundary.}

\section{Flow Equation Approach For Two Chains}
\label{sec:twoChainIntro}
\begin{figure}[t!]
\begin{center}
    \includegraphics[trim={0.3cm 0.1cm 0.1cm 0},clip, width=0.8\textwidth]{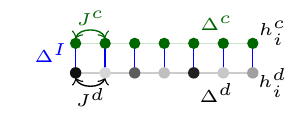}%
\end{center}
\caption{Cartoon of the model described by the Hamiltonian~\eqref{eq:bare}. The MBL sector (bottom wire) acts as a source of disorder to induce localization in the clean component (green sites). The two systems are coupled site by site via inter-chain couplings (blue lines) of strength $\Delta^I$. \label{fig1}}
 \end{figure}
\subsection{The model}
\label{sec:model}

In this section we extend the flow equation method to the system depicted in Fig.~\ref{fig1}.
We consider a system composed of two wires of interacting spinless fermions coupled via an inter-chain density-density interaction of strength $\Delta^I$.
The Hamiltonian of the system reads (cf. Fig.~\ref{fig1})
\begin{eqnarray}
    \label{eq:bare}
    H&=&H^{c}+H^{d}+H^{I} \\ \nonumber
    H^{c} &=& \sum_{ij}J^{c}_{ij}c^{\dagger}_{i}c_{j} + \sum_{ij}\Delta^{c}_{ij}n^c_{i}n_{j}^{c} \\ \nonumber
    H^{d} &=& \sum_{ij}J^{d}_{ij}d^{\dagger}_{i}d_{j} + \sum_{ij}\Delta^{d}_{ij}n^d_{i}n_{j}^{d} +\sum_{k}h_{i}n^{d}_{i} \\ \nonumber
    H^{I} &=& \sum_{ij}\Delta^{I}_{ij}n^{c}_{i}n^{d}_{j}
\end{eqnarray}
where the sums run over $N_{s}$ dirty sites in the Hamiltonian $H^d$ and over $\delta \times N_{s}$ clean sites (with $\delta\geqslant1$) in the Hamiltonian$H^c$. 
The fields, $h_i$, are drawn from a uniform box distribution of variance $W$, i.e. $h_i\in[-W,W]$; for sufficiently large $W$, the chain of fermions, $d_i$, will be in the MBL phase, and will act on the clean fermionic component, $c_i$, as a source of disorder. 
Even though we study a microscopic model that contains inhomogeneities only in the on-site fields, $h_i$, we write couplings in Eq.~\eqref{eq:bare} with a generic dependence on spatial indices to emphasize that, already at the first steps of integration of  the flow equations, couplings can inherit an explicit spatial dependence from the disordered fields.
%

%We study the impact of the disordered environment on the clean system using a variant of the Flow Equation (FE) method, also known as Wegner-Wilson flows (see the broad corpus of literature in~\cite{Kehrein}), tailored to treat MBL systems~\cite{Pekker17, Thomson17}. 
%%
%The approach consists in iteratively diagonalizing the Hamiltonian of the system in real space, for a given realization of disorder, with a series of continuous unitary transformations.
\subsection{Two Chain Ansatz}
Similar to the FE method for single chain MBL phase, the FE method for the two-chain problem aims to construct a unitary transformation, $U(l)=T_l\exp\left(\int\eta(l)dl\right)$, that diagonalizes the Hamiltonian Eq.~\ref{eq:bare}.
    In both cases, an exact calculation would require keeping track of $O(2^{N_s 2})$ matrix elements and is therefore numerically unfeasible.
    As for the MBL phase in the single chain problem, the local nature of the MBL proximity effect allows one to circumvent this issue via an ansatz for the Hamiltonian, $H(l)=U^{\dagger}(l)HU(l)$ at scale $l$ of the unitary transform.
 The ansatz we use for the two-chain problem is $H(l)=H_{0}(l)+V(l)$ where
\begin{eqnarray}
    \label{eq:anzats}
    H_0(l)&=&H^{c}(l)+H^{d}(l)+H^{I}(l) \\ \nonumber
    &H^{c}(l) &= \sum_{ij}\Delta^{c}_{ij}(l):n^c_{i}n_{j}^{c}: +\sum_{k}\bar{h}^c_{k}(l):n^{c}_{k}: \\ \nonumber
    &H^{d}(l) &= \sum_{ij}\Delta^{d}_{ij}(l):n^d_{i}n_{j}^{d}: +\sum_{k}\bar{h}^d_{k}(l):n^{d}_{k}: \\ \nonumber
    &H^{I}(l) &= \sum_{ij}\Delta^{I}_{ij}(l):n^{c}_{i}n^{d}_{j}: \\ \nonumber
    V(l) &=& \sum_{ij}J^{c}_{ij}(l):c^{\dagger}_{i}c_{j}: + \sum_{ij}J^{d}_{ij}(l):d^{\dagger}_{i}d_{j},
\end{eqnarray}
$:A:$ denotes Wick Ordering~\cite{Kehrein}, the fields $\bar{h}^{c(d)}$ are given below in Eqs.~\eqref{eq:effh}, and we will use the convention that the first index in $\Delta^I_{ij}$ refers to the clean chain.
In the limit $l\rightarrow\infty$, $V(l)\rightarrow0$, the fixed-point Hamiltonian, $H(l\to\infty)$, is diagonal in an extensive set of {l-bits \it} localized on both the clean and dirty sites.

As customary for flow equation methods~\cite{Thomson17, Kehrein}, we use Wick-ordered operators, $:A:$, with respect to a reference state $\rho$.
{Wick ordering reduces errors in the truncated Hamiltonian $H(l)$ for the Hilbert space spanned by few particle excitations on top of  the reference state $\rho$~\cite{Kehrein}.
    As done by Thomson {\it et al.} in \cite{Thomson17}, we choose a reference state with zero entanglement between {local degrees of freedom. 
    This extreme locality condition serves a starting point for the FE unitary transformation to capture the  entanglement of the MBL proximity effect.}
}
The state $\rho$ employed is a Boltzmann distribution, $\rho=\frac{1}{Z}e^{-\Theta H_w}$, with inverse temperature $\Theta$, chemical potentials fixing particle densities $\left<n^d\right>$ and $\left<n^c\right>$, and Hamiltonian $H_w=\sum_{i}(h_{i}^{d}-\mu^{d})n_{i}^{d}-\mu^{c}n_{i}^c$.
The choice of  this state allows to easily control  energy density, $\Theta$, and particle density distribution, $\left<n^{d}\right>$.

By Wick-ordering the Hamiltonian at flow time $l=0$, the clean and dirty chains pick up  effective fields, given by
\begin{eqnarray}
    \label{eq:effh}
    \bar{h}_{i}^{d} &= & h_{i}^{d}+2\sum_{j}\Delta_{ij}^{d}\left<n^d_{j}\right>+\sum_{j}\Delta_{ji}^{I}\left<n^c_{j}\right>, \\ \nonumber
    \bar{h}_{i}^{c} &= & 2\sum_{j}\Delta_{ij}^{c}\left<n^c_{j}\right>+\sum_{j}\Delta_{ij}^{I}\left<n^d_{j}\right>;
\end{eqnarray}
where their  distribution  depends  on the dirty chain density, $\left<n^{d}\right>$, the inter-chain coupling $\Delta^{I}$, and the disorder, $W$, in the dirty chain.
From the expressions of the fields in Eq.~\eqref{eq:effh}, it is natural to observe that, if the  dirty chain is sufficiently disordered  and the  inter-chain couplings are sizable,  the clean chain will localize as result of the effective disordered field, $\bar{h}_{i}^{c}$.

Note, the ansatz in~\eqref{eq:anzats} has the notational symmetry 
\begin{eqnarray}\label{eq:DCsym}
c&\leftrightarrow& d \\ \nonumber
\Delta_{ij}^{I} &\leftrightarrow& \Delta_{ji}^{I}.
\end{eqnarray}
By exploiting this symmetry, it is easy to derive flow equations for operators of the dirty chain from those of the clean one, and vice-versa.
We will refer to terms (or equations) produced by such  symmetry transformations using the notion $C\leftrightarrow D$ in the following.

%\subsection{Computing the l-bit Hamiltonian}
 \subsection{Second Invariant and Phase Boundary Analysis}
 \label{sec:secndInv}

From the ansatz in Eq.~\ref{eq:anzats}, we can derive the FE generator $\eta(l)=[H_0(l),V(l)]$.
Then, by matching the truncated terms in the Heisenberg equation of motion, ${dH(l)}/{dl}=[\eta(l),H(l)]$, the {\it truncated flow equations}, a set of first order differential equations for the couplings,
\begin{eqnarray}
    \Gamma=\{\Delta_{ij}(l)^{c(d,I)}, J_{ij}^{c(d)}(l), \bar{h}_{k}^{c(d)}(l)\},
\end{eqnarray}
can be derived 
\begin{eqnarray}\label{eq:truncFlow}
    \frac{d\Gamma}{dl}=\beta(\Gamma),
\end{eqnarray}
where the $\beta$ functions are order three polynomials in the couplings $\Gamma$ and their forms discussed in detail in section~\ref{sec:flowEQMainTxt}.
An {\it l-bit} Hamiltonian, $H(l\rightarrow\infty)$, is retrieved by numerically evolving Eq.~\ref{eq:truncFlow} with initial conditions given by the bare physical couplings, evolved for large $l$.
Deep in the MBL proximity effect phase, these differential equations describe a unitary transform to a diagonal Hamiltonian $H(l\rightarrow\infty)$, and the unitary transformation described by $\eta(l)$, along with the Hamiltonian $H(\infty)$, can be used to predict dynamics of relevant observables~\cite{Thomson17,Kehrein,varma2019a}.

When either chain delocalizes, the {\it l-bit Hamiltonian} ansatz will be an insufficient representation of the effective Hamiltonian, and the couplings $\eta(l)$ and $H(\infty)$ cannot be used to make predictions. 
To detect the breakdown of the MBL proximity effect ansatz, we monitor the extent that the truncated flow equations, Eq.~\ref{eq:truncFlow}, break unitarity.
For this goal, we employ a quantity known as the second invariant, see, for instance, previous work in Refs.~\cite{Thomson17,Monthus_2016}.
The second invariant is the $p=2$ case of a class of many invariants of the FE unitary transformation given by $\text{Tr}[H(l)^{p}]$.
It is particularly easy to calculate for the spin systems and is given by:
\begin{eqnarray}
    &\tr\left[H(l)^2\right] = 
    &\sum_{ij,r=c,d} (J^r_{ij})^2+(\Delta^r_{ij})^2+\Delta^{I}_{ij}+\sum_{k,r=c,d}(\bar{h}^{r}_{k})^2.
\end{eqnarray}
We can then quantify the error made by a given ansatz by computing the change in the second invariant:
\begin{eqnarray}
    \delta I = 2\frac{\tr\left[H(l=\infty)^2\right]-\tr\left[H(l=0)^2\right]}{\tr\left[H(l=\infty)^2\right]+\tr\left[H(l=0)^2\right]}.
\end{eqnarray}
If $\delta I$ is small, then the MBL proximity effect ansatz in Eq.~\ref{eq:anzats}, and the approximations discussed above, represent a reliable description and can be used to compute dynamics and the local conserved l-bits.
On the other hand, when $\delta I$ is large, we have an indication that the ansatz fails and that we cannot use the generator $\eta(l)$ nor the l-bit Hamiltonian $H(\infty)$ to make predictions.

By identifying a threshold for $\delta I$, we can find a {tight bound on the phase boundary for the MBL proximity effect.}
While the choice of threshold is arbitrary, by making it stringently small, one can ensure that below that threshold the MBL proximity effect is properly captured.
On the other hand, if it is above that threshold, we must conclude that 1) the system is delocalized or, 2) it is localized in an operator basis not captured by the ansatz.
If 2) is the case, then, the operator basis must contain either non-local operators or operators capturing stronger correlations.
 In either case, a reasonably chosen threshold should yield an approximate boundary for the MBL proximity effect.
%On the other hand, if it is above that threshold, we can  conclude in support of  one of two following scenarios: 1) the Hamiltonian is not diagonalizable by a local transformation, or 2) the local transformation requires keeping track of more complicated many body operators in the flow equation ansatz.
%Statement 1) implies that the system is delocalized while statement 2) implies the growth of non-trivial correlations and is suggestive that the system is close to delocalization.
%In either case, a reasonably chosen threshold should yield an approximate boundary to the MBL proximity effect phase.

\section{Numerical Results For Equal Length Chains}
\label{sec:numResults}
\subsection{MBL Proximity Effect}
In this section, we present numerical results, for system sizes unattainable with exact diagonalization, that establish the validity of using an l-bit Hamiltonian to describe the MBL proximity effect. .
We study the model introduced in section~\ref{sec:model} for two equal length chains of length $N_s=24$ ($48$ total sites), and numerically solve the differential flow equations, Eq.~\ref{eq:truncFlow}.
For this model, the initial couplings are given as:
\begin{eqnarray}
    \Gamma(l=0)= \\ \Big\{ \nonumber
        \Delta_{ij}^{c(d)}(l=0)&=&\Delta^{c(d)}(\delta_{i,j+1}+\delta_{j,i+1}),\\ \nonumber
        \Delta_{ij}^I(l=0) &=& \Delta^I \delta_{ij} ,\\ \nonumber
        J_{ij}^{c(d)}(l=0)&=&J^{c(d)}(\delta_{i,j+1}+\delta_{j,i+1}),\\ \nonumber
        \bar{h}_{i}^{d}(l=0) &=& h_{i}^{d}+2\sum_{j}\Delta_{ij}^{d}(l=0)\left<n^d_{j}\right>+\sum_{j}\Delta_{ji}^{I}(l=0)\left<n^c_{j}\right>, \\ \nonumber
        \bar{h}_{i}^{c}(l=0) &=& 2\sum_{j}\Delta_{ij}^{c}(l=0)\left<n^c_{j}\right>+\sum_{j}\Delta_{ij}^{I}(l=0)\left<n^d_{j}\right>  \nonumber
    \Big\}
\end{eqnarray}
where $h_{i}^{d}$ is drawn from a box distribution in the interval, $[-W,W]$, and clean and dirty number densities are computed with respect to the Wick ordering reference state, $\left<n^{c(d)}_{k}\right>=\text{Tr}[\rho n^{c(d)}_{k}]$.
We focus on the limit in which the disordered system would be strongly localized and vary the inter-chain coupling, clean chain hopping strength and reference state parameters.
Therefore, we set $W=60$, $\Delta^d=J^{d}=0.1$, and vary the parameters $\Delta^I$, $J^c$, $\left<n^d\right>$($\mu^d$) and $\Theta$.
By setting $\Delta^c=0.1$, we also focus our attention to the limit in which the clean intra-chain coupling is weak.

The exact form of the truncated flow-equations are given in \ref{apx:floweq} and discussed in section~\ref{sec:flowEQMainTxt}.
For a fixed configuration of $h_i^d$, the truncated flow equations are numerically evolved for a sufficiently long flow-time such that 1) the hoppings, $J_{ij}^{c(d)}(l)$, have become sufficiently small, and 2) there is no appreciable change in the flow of any  other coupling.
The evolution is repeated for different random instances of $h_i^d$, and we present the disorder average of the asymptotic ($l\rightarrow\infty$) couplings.

\begin{figure}[t!]
    \includegraphics[width=\textwidth]{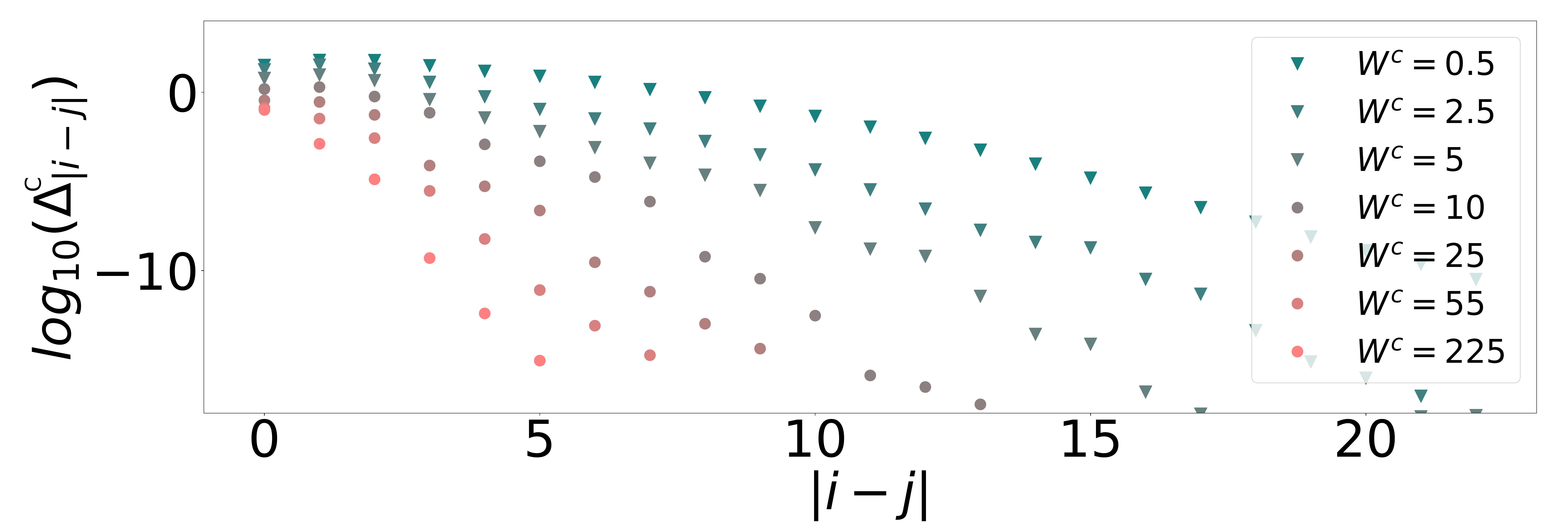} \\
    \includegraphics[width=\textwidth]{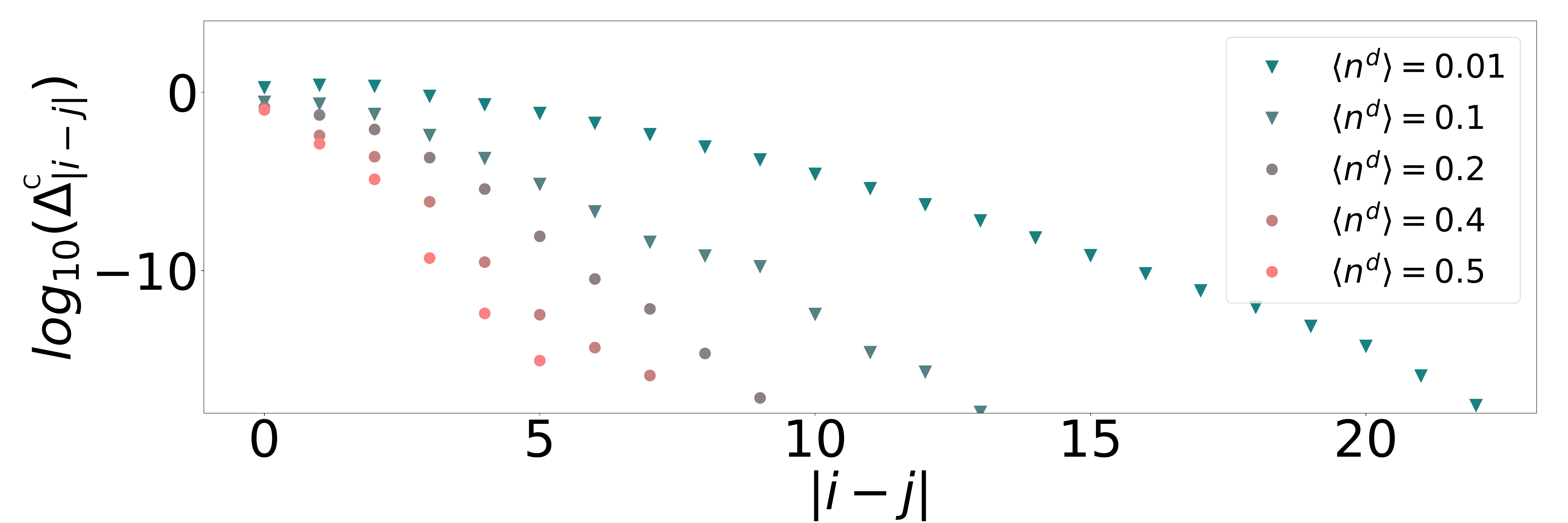}%
    \caption{
    Instances of the MBL proximity effect: the plots show, in logarithmic scale, the spatial decay of the  couplings between  integrals of motion in the clean sector of the system.
    We display results for parameters which yield both a large and small change in the second invariant, and distinguish them using triangle and circle makers respectively.
    The results for parameters that yielded a large change in the second invariant (marked with triangles) do not reflect the true {l-bit} coupling but are displayed to depict how the MBL proximity effect ansatz breaks down.
    The final clean-chain density-density couplings $\Delta^c_{|i-j|}$ depicted here are averaged over  256 disorder realizations.
    In the top panel, we plot how the final density-density couplings depend on $W^c=\Delta^{I}/2J^{c}$ ($J^c$ fixed) while in the bottom panel we plot their dependence on  $\left<n^d\right>$.
    In the top panel $\left<n^d\right>=0.5$ while in the bottom panel $W^c=225$ ($\Delta^I=45$ and $J^c=0.1$).
    The remaining Hamiltonian parameters are $W=60$, $J^d=\Delta^{d}=0.1$,  $\Delta^{c}=J^{c}=0.1$, $\Theta=0.3$, and $\left<n^c\right>=0.1$.
    These  results are not affected by $\left<n^c\right>$ since they are uniformly distributed in the reference state $\rho$ and do not have an impact on the disorder of the effective fields.
}
 \label{fig2}
 %\end{figure}
 \end{figure}

In analogy to a single disordered chain, we define an effective disorder parameter as $W^c=\Delta^I/2J^c$ and work in a limit in which the clean chain is expected to be strongly localized:
$\Delta^I=45$, $J^c=0.1$, $\Theta=0.3$ and $\left<n^d\right>=0.5$ (i.e. $W^c=225$).
We choose such a strong effective disorder to benchmark the method and isolate the effects of varying different parameters.
Solving the numerical flow equations (see \ref{apx:numerics} for details on numerical implementation), we find that
the density-density couplings, $\Delta^{c(d)}_{ij}$, are exponentially suppressed in $|i-j|$, as it occurs in the applications of the Wegner flow to single disordered chains\cite{Pekker17, Thomson17}.
In Fig.~\ref{fig2}a, we show the decay in space of the {disorder-averaged}, asymptotic, density-density couplings, $\Delta^c_{ij}(l\rightarrow\infty)$, on a logarithmic scale, and they illustrate  the onset of an MBL phase in the clean chain.
As discussed below, the change in the second invariant for these parameters is small for the majority of disorder realizations and thus confirms the validity of the MBL proximity effect ansatz employed in this ansatz.

The top panel of Fig.~\ref{fig2} shows that by decreasing the inter-chain coupling, the final density-density couplings between the {\it l-bits} present a slower decay in space suggesting a departure from the MBL proximity phase.
The effective disorder parameter, $W^c=\Delta^I/2J^{c}$, can be used to compare with the disordered Heisenberg chain (a prototype of MBL), which shows a transition at $W/J=4$.
{By considering the second-invariant, we find that the truncation produces minimal error for $W^c\gtrsim10$ and the MBL proximity is well-established.
    Note that while we benchmark the method with $W^{c}=225$, we found the MBL proximity effect to be consistent with a {\it l-bit} ansatz  for a reasonable effective disorder strength of $W^c>10$.
While for $W^c\lesssim10$, the error grows with decreasing $W^c$ and suggests that somewhere in the range  $W^c\lesssim10$ the system undergoes a transition to a delocalized phase.
In this limit, we have found that the final density-density couplings for the dirty-chain, $\tilde{\Delta}^{d}_{ij}$, are still strongly localized while those for the clean-chain are not.
This suggests that the source of truncation error is due to the clean-chain becoming delocalized.}

The bottom panel of Fig.~\ref{fig2}b is  one of the most interesting results of our analysis.
Here, different curves correspond to different fermionic densities of the dirty component in Hamiltonian~\eqref{eq:bare}, with fixed total fermionic density, $\langle n_{tot}\rangle\equiv\langle n^d\rangle+\langle n^c\rangle=0.5$.
This variation of $\langle n^d\rangle$ follows a similar logic to the experiment in Ref.~\cite{imma}, where a complementary situation has been considered (the melting of an MBL phase by coupling to a clean bath).
There, the delocalizing effect of the clean component on the dirty component has been experimentally observed in a mixture  of collisionally coupled ultra-cold bosons in a two-dimensional optical lattice.
Above a certain critical density of bosons,  the clean component  acts as an ergodic bath and destroys the features of the MBL phase in the dirty sector.
Complementary, we find that a critical density of  dirty fermions is required in order for the MBL systems to be sufficiently large to entail localization in the clean component. 
{\blue The analysis of the second invariant identifies that the MBL proximity effect is well-established for  $\left<n^d\right> > 0.25$, and suggests that for some value of $\left<n^d\right>$ less than $0.25$, the clean chain goes through a delocalization transition. It is important to note that we are unable to identify with accuracy the point of transition since our ansatz fails close to it (see also Ref.~\cite{Thomson17}).}

\begin{figure}[t!]
    \centering
    \includegraphics[width=0.5\textwidth]{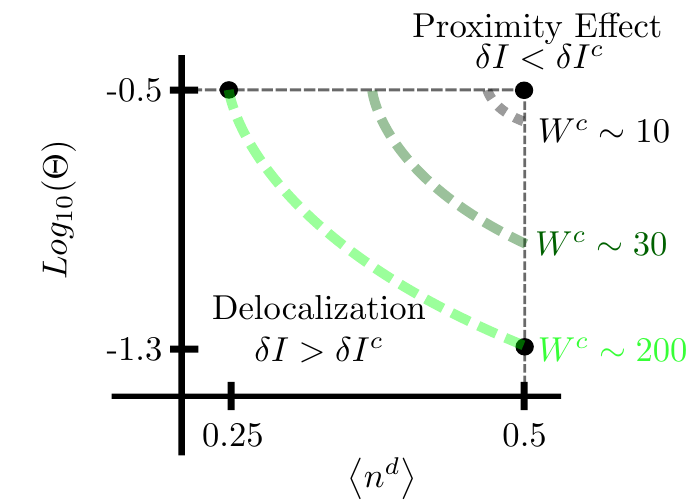}
    \caption{
        Regions of parameter space where the MBL proximity effect is established.
        We focus on an instance of a strongly localized  dirty chain ($W=60$, $\Delta^d=J^{d}=0.1$), and on a clean chain with $\Delta^c=J^c=0.1$.
        The thin, dashed, black lines delimit a square where the parameters $\Theta$ and $\left<n^d\right>$ have been varied in our numerical trials.
        The region of  parameters space in which the MBL Proximity Effect is established is determined by the region where the second invariant is below a specified threshold $\delta I<\delta I^c=0.1$.
        In the figure, we draw   three different thick, curved, dashed lines, corresponding to the values of the parameter $W^c=$10 (gray), 30 (dark green), 200 (bright green).
        These lines mark the values of $\Theta$ and $\left<n^d\right>$ where we expect the second invariant to equal the threshold value $\delta I(W^c, \left<n^d\right>, \Theta)=\delta I^c$, and above which we expect $\delta I(W^c, \left<n^d\right>, \Theta)<\delta I^c$.
        This analysis demonstrates that the MBL Proximity Effect can be observed for the smaller $\left<n_d\right>$ and $\Theta$ when $W^c$ is larger.
        %The interesection of the $W^c=200$ line with the thin, dashed, black line was determine by the numerical trials discussed in the main text.
        %When they intercept the square delimited by the thin, dashed lines, one recovers values the critical values of $\Theta$ and $\left<n^d\right>$ discussed in the text.
        %The intersection of the $W^c=200$ (bright green) line with the thin dashed lines  
      %  For a fixed value of $W^c$, we expect the MBL Proximity Effect to be established ($\delta I<\delta I^c$) in the region above the curved line labeled with the same $W^c$.
}
    \label{fig:phaseDiagram}
\end{figure}

We  have also studied the effect of increasing the clean-chain hopping, $J^c$ and the energy density parameterised by the inverse temperature, $\Theta$ of the reference state $\rho$.
We found that the l-bit ansatz, Eq.~\ref{eq:anzats}, becomes inefficient for large clean chain hopping, $J^c>0.5$, and at large energy densities, $\Theta<0.05$.
In these limits, the clean chain couplings, $\Delta^{c}_{ij}$, begin to delocalize while the dirty chain couplings, $\Delta^{d}_{ij}$, are unaffected.
This dependence of localization on the hopping strength is similar to  a standard MBL system (the system delocalizes at strong hopping), while the dependence on the energy density of the dirty chain is novel.
At low energy density, the dirty chain charge distribution in the reference state, $\left<n_{k}^d\right>$, and, correspondingly, the effective clean disorder fields, $\bar{h}_{k}^c$, are strongly disordered, and the clean chain localizes.
While for high energy density, the reference state has no disorder in the dirty chain densities, and the clean chain delocalizes.
Extrapolating these results, we expect that the localization of the clean chain depends on the disorder of the dirty chain charge distribution.

    We summarize our results in the portrait of Fig.~\ref{fig:phaseDiagram}, which shows the region of the $\Theta$-$\left<n^d\right>$ plane where the change in the second invariant is expected to be smaller than our chosen threshold $\delta I^c=0.1$. 
    In addition to depicting the trends just discussed, it shows  that the dirty chain densities of the reference state must be strongly disorder  to compensate for a weaker inter-chain coupling $\Delta^I$ ($W^c$), in order to induce MBL in the clean sector.

 \subsection{Second Invariant}
 \label{sec:secInvarResults}

\begin{figure}
    \includegraphics[width=0.33\textwidth]{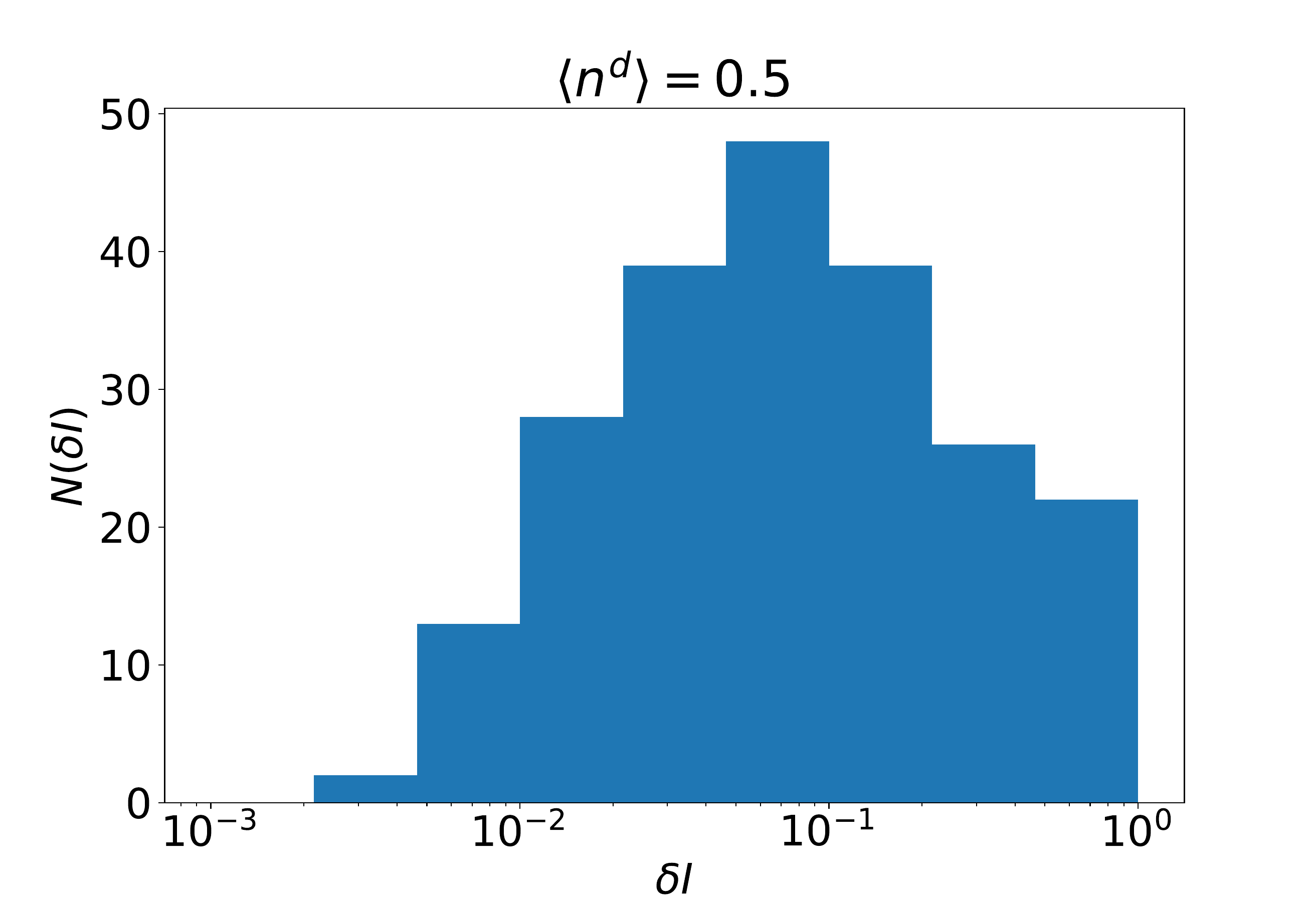}%
    \includegraphics[width=0.33\textwidth]{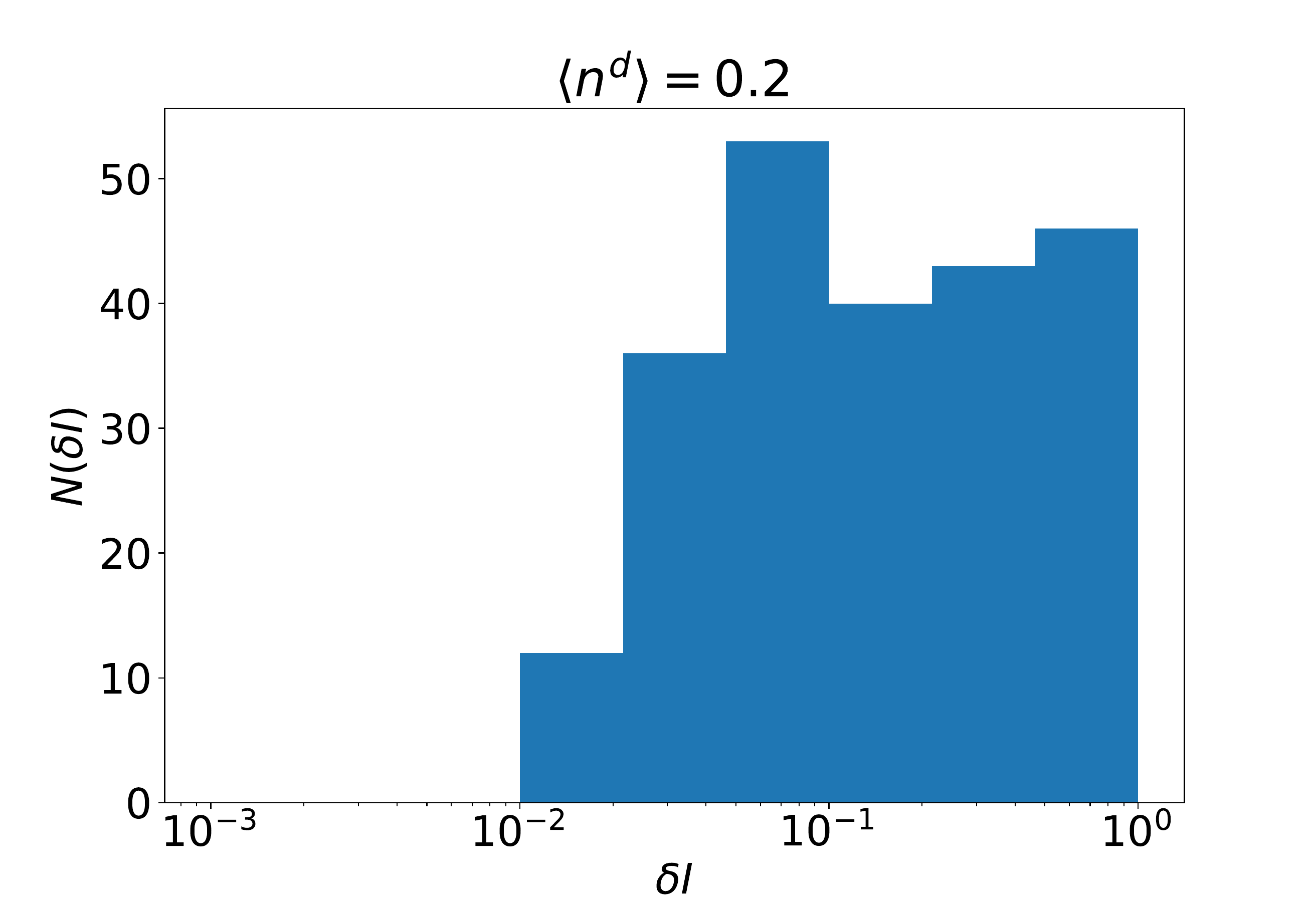}%
    \includegraphics[width=0.33\textwidth]{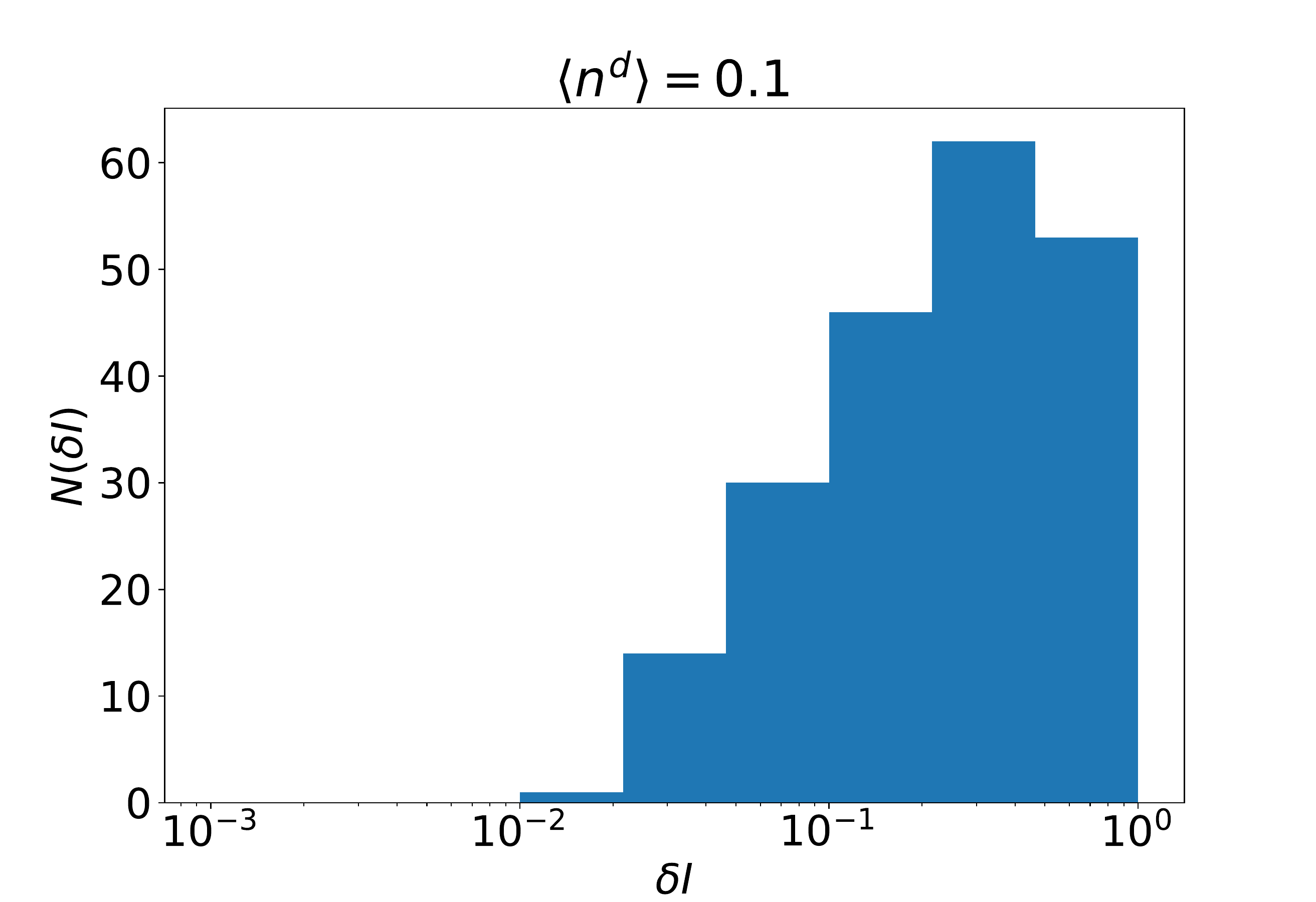}%
    \caption{Histograms of the change in the second invariant as $\left<n^d\right>$ is varied across the transition ($W=60$, $J^d=\Delta^{d}=0.1$,  $\Delta^{c}=J^{c}=0.1$, $\Theta=0.3$, and $\left<n^c\right>=0.1$).}
 \label{fig:secondInvarRef}
 \end{figure}

\begin{figure}
    \includegraphics[width=0.5\textwidth]{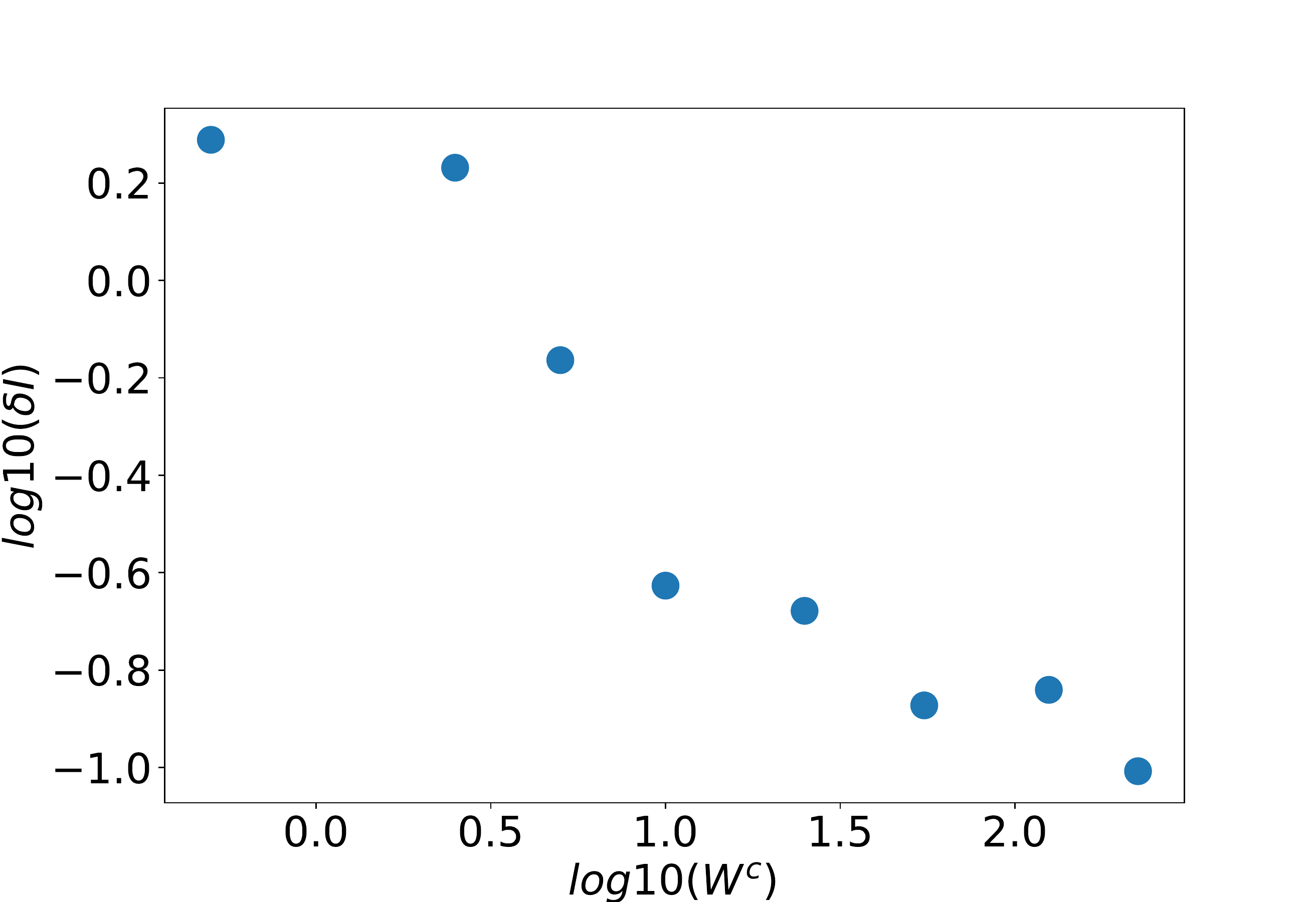}%
    \includegraphics[width=0.5\textwidth]{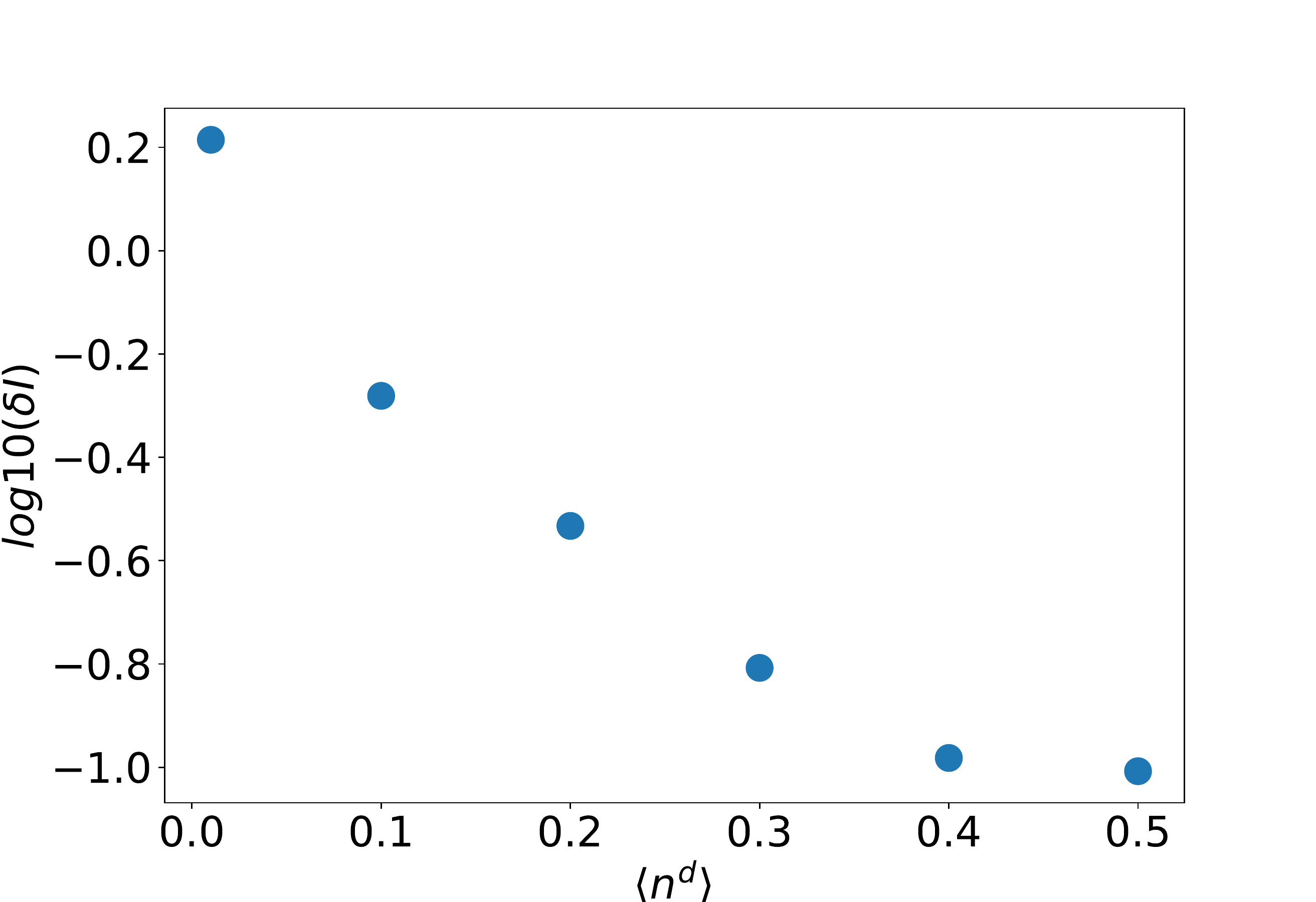}%
    \caption{Median change in the second invariant as a function of $\left<n^{d}\right>$(right) and of $W^c$(left). 
    $W^c$ is plotted on a log scale since it varies over two orders of magnitude. The remaining Hamiltonian parameters are $\Delta^I=45$~($W^c=225$), $J^d=\Delta^{d}=0.1$,  $\Delta^{c}=J^{c}=0.1$, $\Theta=0.3$, and $\left<n^c\right>=0.1$. In the left panel $\left<n^d\right>=0.5$, while, on the right panel, $W^c=225$ ($\Delta^I=45$).}
 \label{fig:secondInvarTrans}
 \end{figure}

 Above we used the second invariant, $\delta I$, to identify when the truncated flow equations preserve the unitarity of the exact Wegner-Wilson flow and to justify the MBL-proximity effect ansatz, Eq.~\ref{eq:anzats}.
Because the flow equation transformation depends on the disorder realization, $\delta I$ varies from sample-to-sample.
The left panel of Fig.~\ref{fig:secondInvarRef} shows the distribution of $\delta I$ for a disorder strength where the MBL-proximity effect ansatz is valid for the majority of disorder realizations, while the right panel shows the distribution for a system where the same ansatz fails for the majority of disorder realizations.
In order to distinguish between these two situations, we can compute the median of $\delta I$ (we don't use the mean because it is {artificially} biased by the few trials with large second invariant weight).
As shown in Fig.~\ref{fig:secondInvarTrans}, the median $\delta I$ shows that the MBL proximity effect ansatz becomes worse for decreasing $\Delta^I$ and $\left<n^D\right>$.
Here we see that for $W^c>10$ and for $\left<n^d\right>>0.25$, the median second invariant is small and relatively unaffected by changes in $W^c$ and $\left<n^d\right>$, demonstrating the validity of the MBL proximity effect ansatz.
While for small $W^c<10$ and small $\left<n^d\right><0.25$, the error made by truncation is large and suggestive of a transition to delocalization somewhere below these values.
The large sample-to-sample variation of the second invariant suggests the presence of regions not captured by the MBL proximity effect ansatz, and future work may attempt to reduce the second invariant for these disorder realizations by improving the ansatz.

\section{Truncated Flow Equations and Truncation Error}
\label{sec:flowEQMainTxt}
\subsection{Truncation Error for the MBL Proximity Effect Ansatz}
\label{sec:AnsatazApprox}
In the previous section we have shown  that the {\it l-bit ansatz}, Eq.~\ref{eq:anzats}, accurately describes the MBL proximity effect phase and that the truncated flow equations, Eq.~\ref{eq:truncFlow}, imply a small  error in approximating the exact flow equation unitary transformation, $U(l)$.
In this section, we analyze the approximations made by the truncation  in Eq.~\ref{eq:anzats}, and we discuss, in section~\ref{sec:contributionsMainText}, the physics of the terms contributing to   the truncated flow equations.
The first type of operators dropped are the $n>2$ body terms such as the three body scattering, $:c^{\dagger}_{i'}c^{\dagger}_{j'}c^{\dagger}_{k'}c_ic_jc_{k}:$.
As long as the integrals of motion do not contain $n>3$ body operators with significant weight, then truncating these terms will not produce significant error in the integrals of motion, FE unitary transformation, or {\it l-bit} Hamiltonian.
This is confirmed by the small second invariant presented in the previous section.
It is important to note that, despite dropping these $n$-body scattering operators, the ansatz does not ignore all $n$-body correlations:  while at scale $l$ the few body terms are not $n>3$ body correlated in the transformed basis, they do contain $n>3$ correlations in the physical basis (i.e. $U(l):c^{\dagger}_{i}(l)c_{j}(l):U^{\dagger}(l)$ contains $n$-body operators).

%The ansatz in Eq.~\ref{eq:anzats} relies on two approximations.
%The first comes from Wick-ordering, and it is equivalent to focus on a specific set of basis states, which reduces  error in dropping four-particle and higher-order operators~\cite{Kehrein}.
%\jm{we have to rephrase this sentence on perturbative expansion or they will kill us again:) let's say it differently, not using perturbaiton theory explicit concepts} The second is a pertubative expansion in the couplings $J^{c(d)}$ and $\Delta^{c(d)}$.
%We will discuss the second in detail here.

In addition to dropping $n>3$ body scattering operators from the {\it l-bit ansatz}, we drop the off-diagonal terms $:n_{k}^cc^{\dagger}_{i}c_{j}:$ and $:c^{\dagger}_{k}c_{l}c^{\dagger}_{i}c_{j}:$, which we will call correlated hopping and full two-body scattering (F.S) respectively.
Including these terms requires keeping track of $O(N_s^3)$ ($O(N_s^4)$ for F.S.) number of couplings and significantly increases the computational resources required.
To identify the error produced by dropping these terms we highlight how they are produced as the flow  evolves.

We identify 7 distinct operators by the 7 sums shown in Eq~\ref{eq:anzats}:
\begin{eqnarray}\label{eq:hamsplit}
    H^{c}(l) &=&  \hat{\Delta}^{c} +\hat{h}^c \\ \nonumber
    H^{d}(l) &=&  \hat{\Delta}^{d} +\hat{h}^d, \\ \nonumber
    H^{I(l)} &=&  \hat{\Delta}^{I} \\ \nonumber
    V(l) &=& \hat{J}^{c} + \hat{J}^{d}
\end{eqnarray}
where, $\hat{h}^{c}=\sum_{k}\bar{h}_{k}^{c}:n^{c}_{k}:$, $\hat{J}^{c}=\sum_{ij}J_{ij}^c(l):c^{\dagger}_{i}c_{j}:$, etc. (see \ref{apx:notation} for explicit forms for the remaining operators).
We then classify contributions to the generator by the type of off-diagonal operator appearing in the commutator: $\eta=[H_{0},J]=\eta_{h}+\eta_{\Delta}+\eta_{I}$ where:
\begin{eqnarray}
    \eta_{h}    &=& [\hat{J}^c,\hat{h}^c]        +  [\hat{J}^d,\hat{h}^d]       \\ \nonumber
    \eta_{\Delta} &=& [\hat{J}^c,\hat{\Delta}^{c}] +  [\hat{J}^d,\hat{\Delta}^{d}]\\ \nonumber
    \eta_{I}    &=& [\hat{J}^c+\hat{J}^{d},\hat{\Delta}^I]. 
\end{eqnarray}
These commutators are computed using rules for Wick ordering~\cite{Kehrein} and yield:
\begin{eqnarray}
    \eta_{h}       &=& \sum_{ij}F_{ij}:c_{i}^{\dagger}c_{j}: +~ C\leftrightarrow D \\ \nonumber
    \eta_{\Delta}  &=& \sum_{ijk}\Gamma^c_{ij|k} :n^c_{k}c^{\dagger}_{i}c_{j}: + F^\Delta_{ij}:c_{i}^{\dagger}c_{j}:+~ C\leftrightarrow D \\ \nonumber
    \eta_{I}      &=& \sum_{ijk}\Gamma^{I}_{ij|k}:n^d_{k}c^{\dagger}_{i}c_{j}: + ~C\leftrightarrow D \\ \nonumber,
\end{eqnarray}
where the coefficients $\Gamma$ and $F$ are given in \ref{apx:generators}.
The form of the generators are either a hopping operator, $:c^{\dagger}_jc_{i}:$, a correlated hopping (C.H) operator, $:n^c_{k}c^{\dagger}_{i}c_{j}:$ or an inter-chain correlated hopping (C.H.I) operator $:n^c_{k}c^{\dagger}_{i}c_{j}:$.
It will be important for quantifying the error implied by our truncation to notice that each of the generators is proportional to $J^{c}_{ij}$ or $J^{d}_{ij}$. In addition, the $\eta_{\Delta}$ generator is also proportional to $\Delta^{c(d)}_{ij}$.

Taking the commutator $[\eta(l),H_{0}(l)+V(l)]$  yields contributions contained both inside and outside the ansatz, $H(l)$, and are summarized in Table.~\ref{tab:terms}.
The operators outside the ansatz are dropped and produce errors proportional to their coefficients.
We expect the majority of these coefficients to be small because we study the MBL proximity effect in a limit that the couplings $\Delta^{c(d)}_{ij}$ and $J^{c(d)}_{ij}$ are initialized as small.
For example, operators appearing in the second row and second and third column appear with coefficients that are proportional to the square of these couplings, and since they are initialized with $\Delta^{c(d)}=J^{c(d)}=0.1$ the error made is $O(0.01)$.

Besides these operators, there are still a few that appear linear in a small coupling and could produce larger error.
For example, $[\eta^{I},\hat{h}^{c}]$ produces an inter-chain correlated hopping operator, $n_k c^{\dagger}_{i}c_{j}$, which has a coefficient proportional to $J^c(\Delta^I)^2$.
While this off-diagonal operator is not small, it is initialized to zero and only affects the diagonal Hamiltonian after commuting with a generator that is also proportional to $J^c$.
Therefore, its effect on the diagonal Hamiltonian will remain small as long as $J^c$ remains small.
This is confirmed by the small change in the second invariant presented above.

\begin{table}[]
    \begin{center}
\begin{tabular}{l|l|l|l|l}
                                          & $\hat{h}^{c(d)}$                                & $\hat{J}^{c(d)}$                                              & $\hat{\Delta}^{c(d)}$                & $\hat{\Delta}^I$ \\ \hline
$\eta_h$                                  & {\color{blue}$J^{c(d)}_{ij} $}                         & {\color{blue}$J^{c(d)}_{ij},h^{c(d)}_i$}                       & C.H.                    & C.H.I      \\
                                          &                                                 &                                                  & {\color{blue}$J^{c(d)}_{ij} $} &           \\ \hline
$\eta_\Delta$     & C.H.                  & {\color{blue}$\Delta^{c(d)}_{i,j}$},C.H., F.S.  &  3P                     & 3P       \\
                                          &                                                 & {\color{blue}$J^{c(d)}_{ij},h^{c(d)}_i$}                       & C.H.                    & C.H.I      \\
                                          &                                                 &                                                  & {$J^{c(d)}_{ij} $} &           \\ \hline
$\eta_I$  & C.H.I                         & {\color{blue}$\Delta_{i,j}^{I}$},C.H.I., F.S.I  &  3P              & 3P         \\
                                          &                                                 & {\color{blue}$J^{d(c)}_{ij},h^{d(c)}_i$}        & C.H.             & C.H.I \\
                                          &                                                 &                                                 &                  & {$J^{c(d)}_{ij} $}\\
\end{tabular}
    \end{center}
    \caption{ This table lists which terms in the commutator $[\eta,H]$ contribute to the beta function $\beta(\Gamma)$ (highlighted in blue) and which are dropped by our ansatz (not highlighted).
    The rows are labeled by the terms in the sum for the generator $\eta=\eta_{h}+\eta_{\Delta}+\eta_{I}$, and the columns are labeled by the terms in the sum for the Hamiltonian, Eq.~\ref{eq:hamsplit}. The notation for the dropped terms is as follows: correlated hopping (C.H.) have a form $n^c_{k}c^{\dagger}_{i}c_{j}$, inter-chain correlated hopping operators C.H.I. have a form $n_{k}^{d}c^{\dagger}_{i}c_{j}$, full scattering terms F.S. have a form $c^{\dagger}_ic^{\dagger}_jc_kc_l$, and 3P terms describing three-body and higher particle scattering. The justification for dropping the contributions to $J^{c(d)}_{ij}$ in the third and forth column is discussed in section~\ref{sec:contributionsMainText}.}
\label{tab:terms}
\end{table}

This completes our analysis of the error produced by the truncation in the Ansatz, Eq.~\ref{eq:anzats}.
In summary, we have discussed how we expect that a small  error will be produced in our truncation scheme, as long as $\Delta^{c(d)}$ and $J^{c(d)}$ are initialized to  small values.
We then referenced results in section~\ref{sec:numResults}, {which demonstrate small truncation error via a small change in the second invariant}, to confirm such expectations.

\subsection{Truncated Flow Equations}
\label{sec:contributionsMainText}
In the previous section, we have sketched the derivation of the FE Heisenberg equation of motion, $\frac{d H}{dl}=[\eta(l),H(l)]$, and discussed the error produced by the truncation of the ansatz.
In this section we focus on operators in $[\eta(l),H(l)]$ that contribute to the ansatz and truncated flow equations~(Eq.~\ref{eq:truncFlow}).
We first focus on the contribution in first row, first column of table~\ref{tab:terms}.
For the clean chain it produces a term:
\begin{eqnarray}
    [\eta_{h},\hat{h}]=&\left[ \left[\hat{J}^c,\hat{h}^c  \right],\hat{h}^c \right]&+\dots= \\ \nonumber
    &-J_{ij}^{c}(\bar{h}^{c}_i-\bar{h}^{c}_{j})^{2}:c^{\dagger}_{i}c_{j}&+\dots
\end{eqnarray}
and therefore contributes to the evolution of $J_{ij}^c$:
\begin{eqnarray}\label{eq:supJ}
    \frac{dJ_{ij}^{c}}{dl}=-J_{ij}^{c}(\bar{h}^{c}_i-\bar{h}^{c}_{j})^{2}+\dots
\end{eqnarray}
This is the primary contribution evolving the off diagonal terms to $0$, and is responsible for the intuitive physics discussed above.
If we ignore the other contributions to ${dJ_{ij}^{c}}/{dl}$ then the evolution of $J_{ij}$ is:
\begin{eqnarray}\label{eq:jijSol1}
    J^c_{ij}(l)=J^c_{ij}(l=0)e^{-(\bar{h}^{c}_i-\bar{h}^{c}_{j})^{2}l}.
\end{eqnarray}
Thus, the stronger the disorder in the effective fields $\bar{h}^{c}_{i}$, the faster the off diagonal terms decay.

In addition to producing terms in the $\beta$ functions that removes the off diagonal couplings $J_{ij}^{c}$, the generator $\eta_h$ renormalizes $\hat{h}^{c(d)}$ and generates off diagonal hoppings $J_{ij}^{c(d)}$ at intermediate scales $l$ of the FE evolution. These terms come from the first row, second column of table.~\ref{tab:terms} and have a characteristic contribution, $[\eta_{h},\hat{J}^{c}]=\left[ \left[\hat{J}^c,\hat{h}^c  \right],\hat{J}^c \right]+\dots$, which produces contributions to the truncated flow equations as:
\begin{eqnarray}\label{eq:growJ}
\frac{d\bar{h}^{c}_{k}}{dl}&= &\sum_{i}2(J^c_{ik})^2(\bar{h}^c_k-\bar{h}^c_{i})  + \dots\\ \nonumber
    \frac{dJ^{c}_{ij}}{dl}&= &-\sum_{k}J^c_{ik}J^c_{kj}(2\bar{h}^c_{k}-\bar{h}^c_{i}-\bar{h}^c_{j})+\dots
\end{eqnarray}
Together with Eq.~\ref{eq:supJ}, Eq.~\ref{eq:growJ} highlights the physics contained in the unitary transformation generated by $\eta_{h}$:  The generator $\eta_h$ is constructed to remove hoppings $J_{ij}^{c(d)}$ that change the energy of the diagonal Hamiltonian, $H_0$, due to the effective fields $\bar{h}^{c(d)}_{k}$.
Eq.~\ref{eq:supJ} shows that the contribution from the commutator $[\eta_h, \hat{h}^c]$ removes off diagonal couplings, 
while Eq.~\ref{eq:growJ} captures new terms produced by the rotation by $\eta_{h}$.

Similar physics occurs for the generators $\eta_{\Delta}$ and $\eta_{I}$, which are constructed to remove hoppings that change energy via the density-density interaction. 
In a strong interacting limit, the exact unitaries produced by these generators will generate a Hamiltonian describing doublon and domain wall propagation\cite{Stein1997}.
If disorder is also strong, these quasi-particle excitation may also localize, realizing a novel MBL of correlated quasi-particles.
Unfortunately, in order to capture these effects, one needs to keep track of computationally demanding correlated hopping operators~\cite{Stein1997} dropped by our ansatz.

While such considerations offer promising prospects for future work, they also have direct consequences for the contributions we include in the truncated flow equations.
{Since the generators $\eta_{\Delta}$ and $\eta_{I}$ transform the hopping operators, $\hat{J}^{c(d)}$, into a set of correlated hopping operators {that commute with  density-density interactions~\cite{Stein1997}}, the truncation above yields a transformation which simply removes the hopping operators without producing the correlated hopping operators they transform into.
If these correlated hopping operators are responsible for delocalization, then removing them would produce an artificial localization.}
To avoid this false localization, we remove the contribution to $\frac{d}{dl}J_{ij}^{c(d)}$ coming from $[\eta_{\Delta},\Delta]$ and $[\eta_{I},\Delta^{I}]$ (respectively, second row, third column; and third row, forth column; of table~\ref{tab:terms}). 
Ignoring these contributions only produces small error for the same reason dropping the correlated hopping operators produces small error: the error in $J_{ij}$ is proportional to $J^{c(d)}_{ij}$ but its contribution to the {\it l-bit} Hamiltonian is $(J^{c(d)}_{ij})^2$.  The small error is numerically confirmed by a small second invariant as discussed above.

The remaining contributions from $\eta_{\Delta}$ and $\eta_{I}$ are the ones in the second column of table~\ref{tab:terms} and describe delocalization processes produced by density-density interactions.
%This rate of decay is in competition with the rate of growth of the density-density couplings.
%The evolution of these couplings are produced by terms in the second row, second column of table~\ref{tab:terms}:
A characteristic contribution is:
\begin{eqnarray}
    [[\hat{J}^{c},\hat{\Delta}^{c}],\hat{J}^{c}],
\end{eqnarray}
which produce a contribution to the evolution of $\Delta^c_{ij}$ as:
\begin{eqnarray}\label{eq:ddijdl}
     \frac{d\Delta_{ij}^{c}}{dl}&=&2\sum_{k\neq i,j l=i,j}J^{2}_{lk}(\Delta^c_{ij}-\Delta^c_{kl'}).
\end{eqnarray}

This contribution captures how the truncated flow equations break the unitary character of the FE transform in a delocalized limit. 
When disorder is small, $J_{ij}$ remains finite longer during the flow equation evolution and $\Delta_{ij}^{c}$ has a longer time to grow according to the contribution in Eq.~\ref{eq:ddijdl}.
This growth produces larger truncation error because, as discussed in the previous section, truncation error is only small when $\Delta_{ij}^{c}$ is small.  
This concludes our analysis of the physical content of the contributions to the truncated flow equations.  The full set of truncated flow equations used in our numerics  are reported in \ref{apx:floweq}.

 \section{Engineering the geometry of the inter-chain couplings}
 \label{sec:novelGeometry}
\begin{figure}[t!]
    \includegraphics[width=\textwidth]{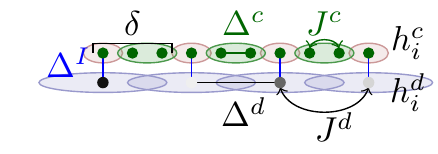}%
    \caption{The  dirty chain couples to the clean chain every $\delta=3$ sites.  
              The emergent integrals of motion are illustrated  with different colors: $n^{d}_{k}$ (blue), $n^{c}_{f,r=0}$(red) and $N_{f}$ (green).
 \label{fig:chainDiag}} 
 \end{figure}
 {
We now discuss novel effects arising by tuning the coupling geometry.
In the geometry of Fig.~\ref{fig:chainDiag}  each fermion of the dirty chain is coupled,  every $\delta$ sites, to a fermion of   the clean chain.
This new geometry can still be studied using analogous flow equations to those employed above.
Since the clean chain is $\delta$ times longer than the dirty chain, we can label the dirty chain with $f=0\dots N_{s}-1$, and conveniently reference the sites of the clean chain ($k=0\dots N_{s}\delta-1$) with $r$, using $k=f\delta+r$.
$f$ labels the dirty sites, and $r=0\dots\delta-1$ is the number of sites away from the coupled site.
We can now explicitly write the initial inter-chain coupling as $\Delta_{f,r,f'}=\Delta^{I}\delta_{f,f'}\delta_{r,0}$.
This leads to an initial clean-chain effective field of $\bar{h}^{c}_{f,r}=\Delta^{I}\langle n^{d}_{f}\rangle\delta_{r,0}$.
}

\begin{figure}[t!]
    \includegraphics[width=0.5\textwidth]{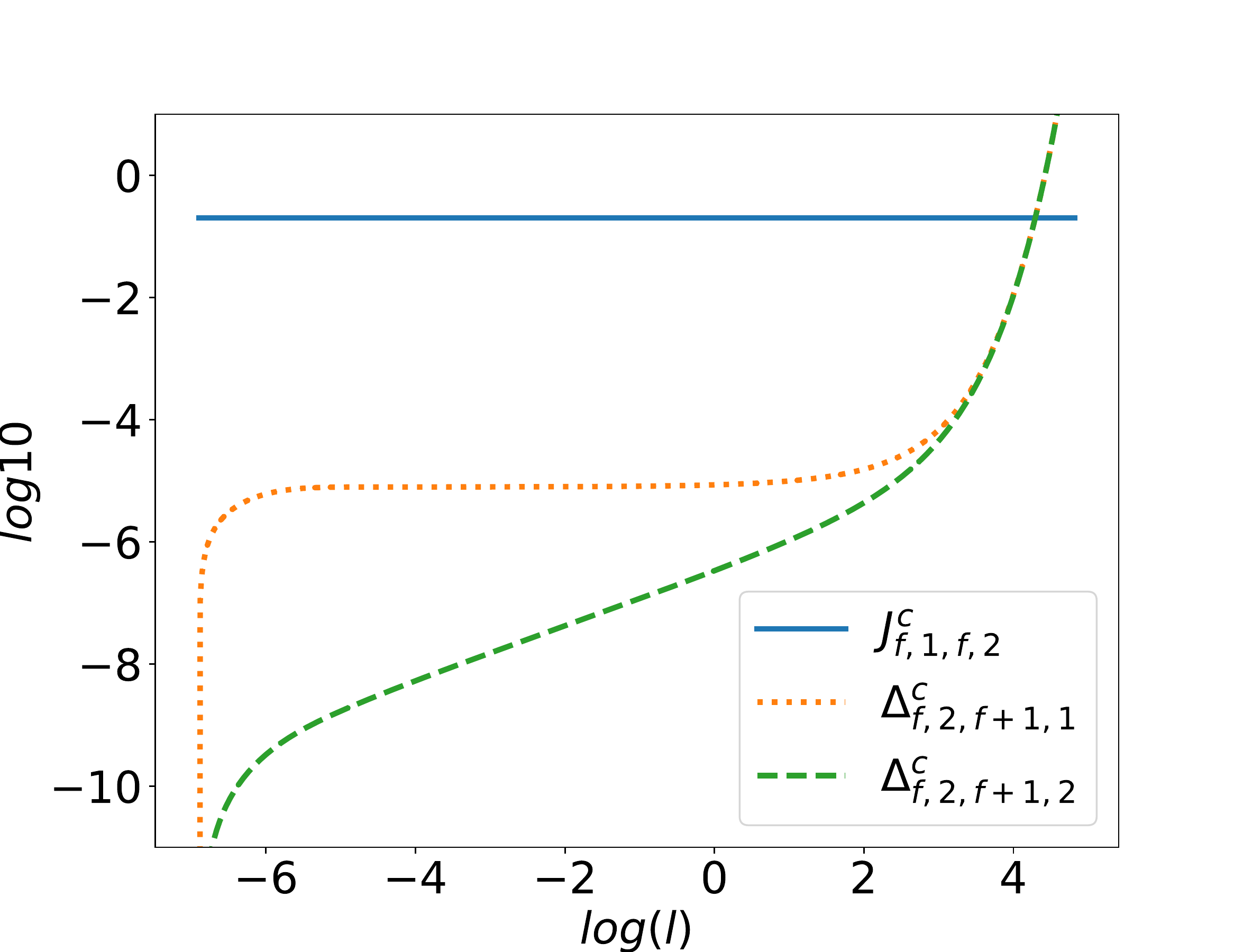}%
    \includegraphics[width=0.5\textwidth]{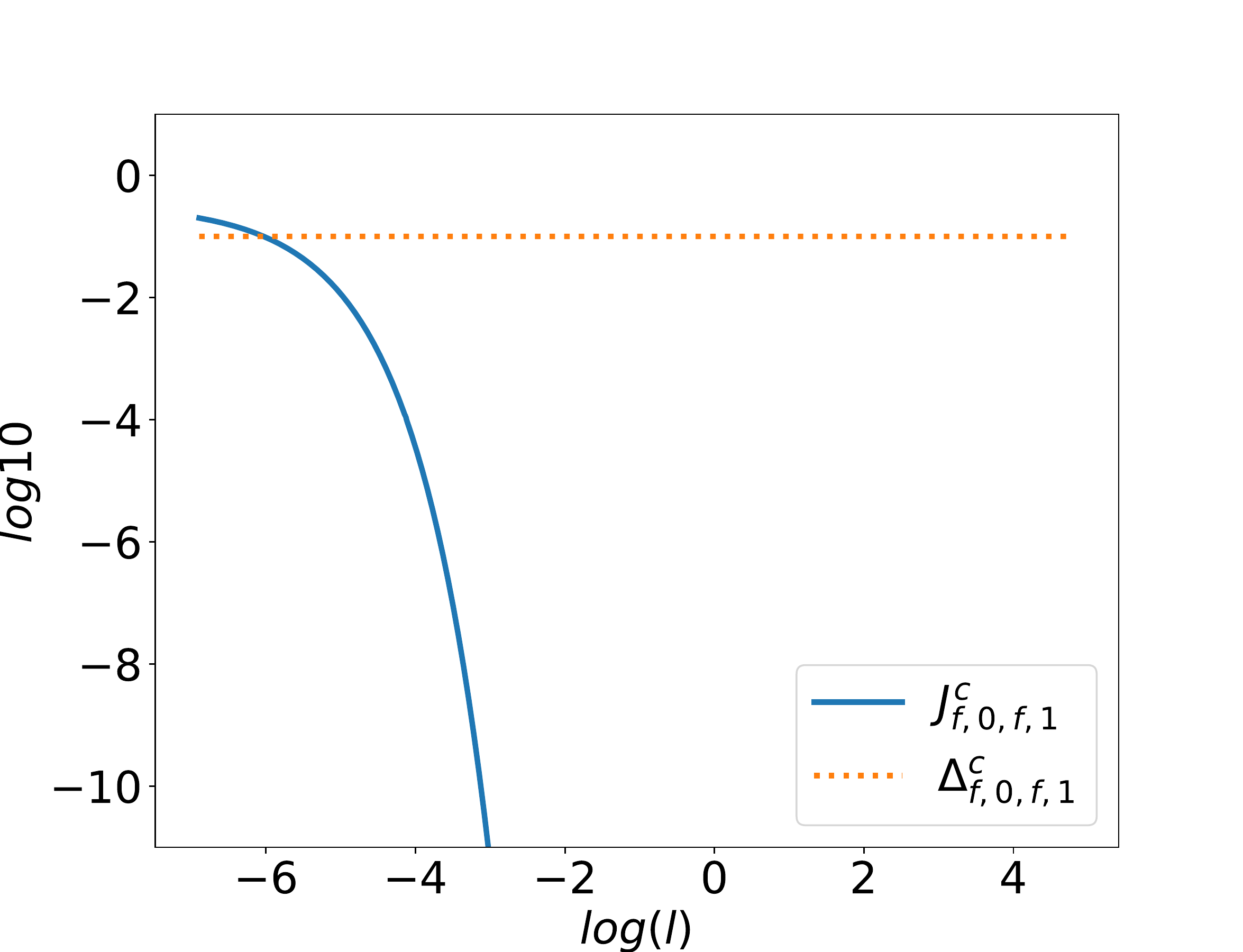}%
    \caption{
        The flow of $J_{ij}^{c}$ and $\Delta_{ij}^{c}$ for the geometry depicted in Fig.~\ref{fig:chainDiag}.
        The left panel shows the flow of couplings on the clean-chain sites that are not coupled to the dirty chain.
        It shows an unsuppressed hopping and diverging density-density coupling at long flow time $l$.
        The right panel shows the flow of couplings on the clean chain sites that involve a site coupled to the dirty chain.
        It shows that the hopping onto the coupled site, $r=0$ (for any $f$), are suppressed and the density-density coupling involving a coupled site, remains constant instead.
        This calculation has been performed  using an unmodified generator $\eta=[H_0,V]$; in order to remove the divergences in $\Delta^c_{f,r,f',r'}$, we modify the generator to $\eta=[H_0,V']$, with $V'$ given in Eq.~\ref{eq:newV}.
    }
    \label{fig:flowd3}
 \end{figure}

{
With this important modifications, we can straightforwardly evolve the couplings using the same truncated flow equations discussed in the previous sections.
We show evolution of few of them  in Fig.~\ref{fig:flowd3}.
The right panel of Fig.~\ref{fig:flowd3} shows the suppression of the hopping between a coupled site $f,r=0$ and an uncoupled site $f,r=1$,
while the left panel of Fig.~\ref{fig:flowd3} shows the hopping between two uncoupled sites, $f,r=1$ and $f'=f,r=2$, remaining constant.
This is consistent with the expectations given by Eq.~\ref{eq:jijSol1}: for a particle to hop on to a coupled site its energy must change by $(\bar{h}^c_i-\bar{h}^c_j)\approx\Delta^{I}\left<n^{d}\right>$, while such a change of energy is not required for a particle hopping between two uncoupled sites.
}

{
With the hopping between uncoupled sites remaining constant, Eq.~\ref{eq:ddijdl} predicts the divergence of the associated density-density couplings.
This is depicted in the left panel of Fig.~\ref{fig:flowd3} and explains the failure of the MBL proximity effect ansatz.
Instead of modifying the ansatz, we propose to modify the generator, $\eta(l)$ of the unitary transformation.
We define a modified generator $\eta'=[H_0,V']$, where we choose $V'$ to only include hoppings  to coupled sites:
\begin{eqnarray}\label{eq:newV}
    V'(l)=\sum_{f,f'}J_{f,f'}^{d}d^{\dagger}_fd_{f'}+\sum_{f}J_{f,0,f,1}^{c}(c^{\dagger}_{f,0}c_{f,1}+h.c.)+J^c_{f,0,f-1,\delta-1}(c^{\dagger}_{f,0}c_{f-1,\delta-1}+h.c.)
\end{eqnarray}
}

{
    {Using such a generator, one can employ the same ansatz as above, but the transformation now results in a novel fixed point Hamiltonian describing  transport between uncoupled sites and conserved charges on coupled and dirty sites ($n^c_{f,r=0}$ and $n^{d}_{f}$ respectively).}
In addition, the new generator produces a next-nearest neighbor hopping across the coupled site (i.e $J_{f,r=\delta-1,f'=f+1,r=1}$).
In the proceeding section we derive this hopping rate  as $\frac{1}{\tau_{n}}={[J^{c}(l=0)]^2}/{\bar{h}^{c}(l=0)}$, which in the limit of strong inter-chain coupling, $\Delta^I$, is smaller than  the other timescales in the system.
In this limit, relaxation occurs in two steps: first, on times scales shorter then $\tau_n$, transport is blocked by the coupled sites, and second, on times scales longer then $\tau_n$, charge is allowed to diffuse across the coupled sites.
In the first step, when  $\tau\lesssim\tau_{n}$,  the system relaxes to a state in which the charge on the bunches of uncoupled sites, $N_f=\sum_{r=1}^\delta n^{c}_{f,r}$, is conserved.
While on longer times, charge on the uncoupled sites can fully relax via unconstrained transport.
In the following two sections, we further investigate this novel behavior by first, in section~\ref{sec:smallBarrier}, deriving the time, $\tau_{n}$, separating the two relaxation steps, and second, in section~\ref{sec:effHam}, investigating and deriving the Hamiltonian that governs short time relaxation.
}

\subsection{Separation of Time Scales }
\label{sec:smallBarrier}

To derive an estimate of the next-nearest neighbor hopping rate, we first assume $\Delta\ll W$.
This guarantees that the flow of the dirty chain reaches a steady state before there are significant changes in the clean one.
We can then treat the clean chain as a single chain with an effective field $\bar{h}_{f,r}$.
We  write the new generator as
\begin{eqnarray}
    \eta'=\sum_{f}\eta_{f}
\end{eqnarray}
where
\begin{eqnarray}
\eta_{f}=-J\bar{h}_{f,0}(c^{\dagger}_{f-1,\delta-1}c_{f,0}-h.c) + J\bar{h}_{f,0}(c^{\dagger}_{f,0}c_{f,1} - h.c),
\end{eqnarray}
with $J$  the strength of the hopping on to the coupled site.
The first term in $\eta_f$ will suppress hopping between the coupled site and its left neighbor, while the second term will enforce the same on the right neighbor.
Since $[\eta_{f},\eta_{f'}]=0$ for $\delta>2$, we can focus on a single coupled site and its neighbor.

We will label the coupled site with $0$ and its left and right neighbor sites with $-1$ and $+1$.
The hopping and effective field couplings will then flow as
\begin{eqnarray}
    \frac{d\bar{h}_{\pm 1}}{dl}&=&-2J^{2}\bar{h}_{0}, \\ \nonumber
    \frac{d\bar{h}_{0}}{dl}&=&2J^{2}\bar{h}_{+1} + 2J^{2}\bar{h}_{-1}, \\ \nonumber
    \frac{dJ}{dl} &=& -J\bar{h}_{0}^2, \\ \nonumber
    \frac{dJ_2}{dl} &=& 2J^{2}\bar{h}_{0},
\end{eqnarray}
where $J_2$ is the magnitude of the next-nearest neighbor hopping, $J_2(l) = J_{f,r=\delta-1,f'=f+1,r=1}(l)$.

The flow of these couplings do not depend on the flow of the density-density coupling and can thus be solved independently.
We use the assumption that $J\ll\bar{h}$ and note that the flow of $J$ is much faster than the flow of the other couplings.
Thus, assuming $\bar{h}$  constant, we can approximate the flow of $J(l)$ as
\begin{eqnarray}
    J(l)=J(l=0)e^{-\bar{h}_{0}^{2}l}.
\end{eqnarray}
Approximating $\bar{h}_{0}$ as constant,  we find
\begin{eqnarray}
    J_{2}(l)=-\frac{J^{2}}{\bar{h}_0}(1-e^{-2\bar{h}_0^{2}l}).
\end{eqnarray}
Thus, $\tau_{n}=\frac{\bar{h}_0}{J^{2}}$ is the characteristic  time when relaxation  crosses over to full transport and eventually to thermalization.
A meaningful separation of time scales therefore requires $\bar{h}_{0}\gg J^{2}$.
In the following section we will discuss the form of the effective Hamiltonian describing the first stage of relaxation.

\subsection{Effective Hamiltonian at intermediate times: $\tau\lesssim \tau_{n}$}
\label{sec:effHam}
{
As discussed above, relaxation in the novel geometry with large inter-chain coupling, occurs in two stages: first, during intermediate times, the model relaxes to a state in which the clean-charge distribution on the uncoupled clusters is approximately conserved, while, on longer times,  the clean-charge relaxes to a homogeneous distribution.
The Hamiltonian describing the first relaxation process is obtained by dropping the next nearest neighbor hoppings from the Hamiltonian,  $H(l\rightarrow\infty)$.
This Hamiltonian, has 3 types of conserved charges as depicted in Fig.~\ref{fig:chainDiag}: the first type, $n_k^d$, are the conserved charges on the dirty chain, the second type $n^{c}_{f,r=0}$ are the conserved charge on the coupled site and $N_{f}=\sum_{r=1}^{\delta}n_{f,r}^{c}$ is the total conserved charge on an uncoupled cluster.
For $\delta>2$ these charges do not determine the dynamics of the charge distribution within an uncoupled cluster, and we must consider the interplay between the intra-cluster tunneling and inter-cluster density-density coupling.
}

There are two possibilities for such interplay: the density-density coupling between two neighboring sets of uncoupled sites is smaller than $J_2$, or it is larger:

\begin{itemize}
\item In the first case, the  density-density coupling can be accurately dropped from the intermediate time effective Hamiltonian.
This leads to each set of uncoupled sites, labeled by $f$, evolving completely independently on intermediate times.
The dynamics can be described  as the evolution of an effective spin, $\vec{L}_f=\{L_x, L_y, L_z\}$, of size 
\begin{eqnarray}
    \left|\vec{L}_{f}\right|=\frac{1}{2}\begin{pmatrix} \delta -1 \\ N_f \end{pmatrix}+\frac{1}{2}.
\end{eqnarray}

The local map between the $N_f$ fermions on $\delta-1$ sites and the spin can be performed by identifying the basis states labeled by the eigenvalues of $n_{f,r\neq0}^{c}$ with the basis states labeled by the eigenvalues of $L_f^z$.
Operators that are polynomial in the densities will then be mapped to operators that are polynomial in $L_z$.
The remaining terms in the Hamiltonian describe tunneling within a set of uncoupled sites with all the same $f$.
They describe transition between the $L_f^z$ basis states and are thus described by polynomials in $L_f^x$ and $L_f^y$.

\item In the second case, {when the density-density interaction between the uncoupled cluster is relevant, the local emergent spins will be coupled.}
Since the hopping operators at a site $f$ commute with those at a site $f'$, a Jordan-Wigner string is not required to correctly reproduce spin statistics, and the coupled Hamiltonian can be written as:
\begin{eqnarray}
    H(\{\bar{n}_f^{d}\},\{\bar{n}_{f,r=0}^{c}\}, \{\bar{N_{f}}\} )= \sum_{ff'}F(L^{z}_{f},L_{f'}^{z}) + \\ \nonumber
\sum_{f}R_f(L_{f}^{x},L_{f}^{z}, L_f^y),     \label{eq:effH}
\end{eqnarray}
where the function $F$ depends on the intra-chain coupling, $\Delta^c$, and the function $R$ depends on $h^{c}_{k},J^{c}_{ij},\Delta^{I}_{ij}$, and $\Delta^{c}_{ij}$.
In general, if the dirty chain or coupled sites have a disordered distribution of charges, the local operators, $R_f$, in the Hamiltonian will be disordered too. 
The issue of whether the system is fully localized on intermediate times, will then depend on any integrability present in this intermediate time Hamiltonian, or on the impact of disorder on $R_f$.
\end{itemize}
{In the first case, the intermediate time Hamiltonian can be diagonalized by independently diagonalizing the Hamiltonian of effective spins $L_f$.
    In the second case, when the spins are coupled, further analysis is required to explore the dynamics at intermediate times and will be the subject of Sec.~\ref{sec:eme}.
}

%\begin{figure}[t!]
    %\includegraphics[width=\textwidth]{Figures/chain}%
     %\caption{The portrait shows the dirty chain coupling to the clean chain every $\delta=3$ sites.
              %The emergent integrals of motion $n^{d}_{k}$ (blue), $n^{c}_{f,r=0}$(red) and $N_{f}$ (green) are also depicted.
              %This picture requires a sufficiently large $\Delta^I$ and $\left<n^d\right>$, but does not require disorder in the effective field $\bar{h}^c$.
              %When the clean-chain effective field is disordered, the remaining degrees of freedom contained in the uncoupled sites can localize.
 %\label{fig:chainDiag}} 
 %\end{figure}

 \subsection{Density-Density interactions between uncoupled clusters}
 To determine if the effective spins $\vec{L}_{f}$ are coupled or not, we compute the magnitude of the density-density interaction between two uncoupled clusters.
 We  focus again on one coupled site, labeled by $r=0$, and its neighboring sites, labeled by $r=\pm1$ (for any $f$).
 The flow equation equations for the density-density couplings then becomes
 \begin{eqnarray}
     \label{eq:deltaflowsimp}
     \frac{d\Delta^c_{-1,1}}{dl}&=&2J^{2}(\Delta^c_{-1,1}-\Delta^c_{0,1})+ 2J^{2}(\Delta^c_{-1,1}-\Delta^c_{0,-1}), \\ \nonumber
     \frac{d\Delta^c_{0,1}}{dl} &=&-2J^{2}(\Delta^c_{-1,1}-\Delta^c_{0,1}), \\ \nonumber
     \frac{d\Delta^c_{-1,0}}{dl}&=&-2J^{2}(\Delta^c_{-1,1}-\Delta^c_{0,-1}). 
 \end{eqnarray}
 These coupled differential equations describe a rotation in a three-dimensional space at an instantaneous rate $2J(l)^{2}$. Given that $\Delta^c_{-1,1}(l=0)=0$, the system~\eqref{eq:deltaflowsimp} can be solved and yields
 \begin{eqnarray}
     \Delta^c_{-1,1}(l)=\Delta^c_{0,1}(l=0)\left[1-e^{\int_{0}^{l}dl'2J^{2}(l')}\right],
 \end{eqnarray}
 where:
 \begin{eqnarray}
     \int_{0}^{l}dl'2J^{2}(l') = \frac{J^{2}(l=0)}{\bar{h}_{0}^{2}(l=0)}(1-e^{-2h_{0}^2l}).
 \end{eqnarray}
 Therefore, the amplitude of the rotation in such three-dimensional parameter space is small in $J^2/\bar{h}_{0}^2$.\\

 We are now in place to discuss which of the  two possibilities discussed in the previous section is realized.
 If $J_{2}(l=\infty)\ll\Delta_{-1,1}(l=\infty)$, then an interacting Hamiltonian  describes the intermediate time dynamics while, if the inequality is not satisfied, a non-interacting spin chain will describes the intermediate time dynamics. Given the assumption $J\ll h$, this inequality simplifies to $h\ll \Delta$.
 Thus, for the approximation made in ansatz Hamiltonian above, we must choose  $h>\Delta$ and conclude that the intermediate time Hamiltonian describes a set of independently evolving spins.

 Alternatively, we could assume the bare Hamiltonian has a  next-nearest neighbor coupling of the order $\Delta_{-1,1}(l=0)\approx \Delta_{0,1}<h$.
 In this case the rotation in $\Delta_{ij}^c$ space, described by Eq.~\ref{eq:deltaflowsimp}, would still be of a small angle, but away from an initial vector with $\Delta_{-1,1}(l=0)$ already greater than $J_{2}(l=\infty)$.
 Intermediate time dynamics would then be described by a set of coupled emergent spins of size $\left|\vec{L}_{f}\right|$.

\subsection{Explicit form of the Hamiltonian for $\delta=3$}
\label{sec:eme}
As an example, we now can consider the $\delta=3$ case in which there are two uncoupled sites for each dirty site $f$, and discuss the effective Hamiltonian governing the intermediate  time dynamics.
The local Hilbert space for these two sites is 4 dimensional and the basis vectors can be labeled by the different ways in which 2 sites may be occupied with particles (the label $'1'$ indicates an occupied site)
\begin{eqnarray}
\begin{Bmatrix}
    \left|00\right>, &
    \left|01\right>, &
    \left|10\right>, &
    \left|11\right>.
\end{Bmatrix},
\end{eqnarray}
 The local Hamiltonian on these sites reflects the block diagonal structure enforced by the conserved charges:
\begin{eqnarray}
    \begin{bmatrix}
        0 & 0 & 0 & 0 \\
        0 & \tilde{\Delta}^L & J^{c}_{f2,f1}(l) & 0 \\
        0 & J^{c}_{f1,f2}(l) & \tilde{\Delta}^{R} & 0 \\
        0 & 0 & 0 & \tilde{\Delta}^{R+L} \\
    \end{bmatrix},
    \label{eq:locH}
\end{eqnarray}
where $\tilde{\Delta}^{L}$,$\tilde{\Delta}^{R}$, and $\tilde{\Delta}^{R+L}$ are functions linear in the operators $n^{d}_{i}$ and $n^{c}_{f'\neq f}$ and depend on the intra and inter-chain couplings, and fields $h^{c}$, at the flow time $l=\infty$.
For $\delta=3$ the conserved charge $N_f$ has eigenvalues $0,1,$ and $2$ that correspond to the three blocks in Eq.~\ref{eq:locH}.
This block structure can be represented by two trivial spin-zero subspaces and one spin-half subspace.

We consider the case that $N_f=1$ for each $f$, so that the local Hilbert space for the block of interest will be spin-half. The mapping to spin-halves can be preformed via
\begin{eqnarray}\label{map}
    L^{z}_{f} &= & \frac{\hat{n}_{f,1}-\hat{n}_{f,2}}{2}  \\ \nonumber
    L^{x}_{f} &= & \frac{c^{\dagger}_{f,1}c_{f,2} + h.c}{2}, \\ \nonumber
\end{eqnarray}
and the constraint $ \frac{1}{2} = \frac{\hat{n}_{f,1}+\hat{n}_{f,2}}{2}$.

We write down the Hamiltonian at the flow time $l=\infty$ as follows:
\begin{eqnarray}
    H&=& \sum H_{f} + \sum_{f,f'}\sum_{r,r'=1,2}\Delta^c_{f,r,f',r'}n^{c}_{f,r}n^{c}_{f',r'} \\ \nonumber
    H_{f}&=& \sum_{i} \xi_{f,i}n_{f,i} + J_{f}^{un}(l) [c^{\dagger}_{f,1}c_{f,2} + c^{\dagger}_{f,2}c_{f,1}] \\ \nonumber
    &&+ \Delta^{c}_{f,1,f,2}n^{c}_{f,2}n^{c}_{f,1}, \\ \nonumber
\end{eqnarray}
where $\xi_{f,i}$ is an effective field that depends on the bare fields at flow time $l$, the couplings $\Delta^{c}_{ij}$ and $\Delta^{I}_{ij}$, and the eigenvalues of the conserved charges, $\bar{n}^{d}_{f}$ and $\bar{n}^{c}_{f,r=0}$:
\begin{eqnarray}
    \xi_{f,i}=\bar{h}^c_{f,i}(l)+\sum_{f}\Delta^{I}_{f,i,f'}(l)\bar{n}_{f'}^{d}+\sum_{f}\Delta^{c}_{f,i,f',0}(l)\bar{n}_{f',0}^{c}. \nonumber
\end{eqnarray}
Applying the mapping~\eqref{map} we get the spin Hamiltonian:
\begin{eqnarray}\label{eq:effH2}
    H = \sum_{f}h^{z}_{f}L^{z}_{f}+h^{x}_{f}L^{x}_{f} + \sum_{ff'}\Omega_{f,f'}L^{z}_{f}L^{z}_{f'} + C
\end{eqnarray}
with 
\begin{eqnarray}\label{eq:hz}
    h^{x}_{f} &= & 2J^{un}_{f}(l)  \\ \nonumber
    h^{z}_{f}&= &\xi_{f,1}-\xi_{f,2} \\ \nonumber
\Omega_{f,f'}&= &\Delta^{C}_{f,1,f',1} + \Delta^{C}_{f,2,f',2} - \Delta^{C}_{f,1,f',2} - \Delta^{C}_{f,2,f',1}.
\end{eqnarray}

Here, we explicitly see how the spins are coupled by the next-nearest neighbor density-density couplings.
Thus, if the local spins are coupled at a strength less than the next-nearest neighbor hopping, $|\Omega_{f,f'}|<J_{2}$, the intermediate time dynamics describes independent spins rotating around an axis in the $x-z$ plane.
While, if $|\Omega_{f,f'}|>J_{2}$,  we have to consider the interacting spin problem to understand the intermediate time dynamics.

If there is no disorder in the dirty and coupled site charge distributions, the $z$ component of the local field, $h_{f}^{z}$ will be null and the translationally-invariant emergent spin-model will be a transverse field Ising model.
This Hamiltonian is integrable via the Jordan-Wigner transformation:
\begin{eqnarray}
    L^{x}_f&\rightarrow& n^{a}_{f}-1/2 \\ \nonumber
    L^{z}_{f}L^{z}_{f+1}& \rightarrow&(a^{\dagger}_{f}-a_{f})(a_{f+1}+a^{\dagger}_{f+1}),
\end{eqnarray}
which produces an exactly solvable single particle Hamiltonian in  Jordan-Wigner fermions.
Taking $\Omega_{f,f'}=\Omega\delta_{f',f+1}$, this single particle Hamiltonian is given as
\begin{eqnarray}
    \sum_{f}h^{x}n_{f}^{a}+\Omega(a^{\dagger}_{f}a_{f+1}+h.c)+\Omega(a^{\dagger}_{f}a^{\dagger}_{f+1}+h.c),
\end{eqnarray}
which can be brought in diagonal form $\sum_q \omega_{q}n_q$ in momentum space via a Bogolyubov rotation, where $n_q$ is the occupation of the mode $q$ and $\omega_{q}=\sqrt{1+2\frac{\Omega}{h}\cos(q)+\frac{\Omega^2}{h^2}}$.
We therefore, in addition to the local conserved charges, $n_{f,0}^{c}$, $n_{f}^d$, $N_{F}$, have the conserved momentum space modes $n_q$.
The non ergodic behavior during intermediate times after the initial relaxation period and before $\tau_{n}$ will display a mixture of local conserved charges, and extended conserved charges, $n_q$.

{If there is disorder in the dirty and coupled site charge distributions, the $z$-components of the local field, $h_z$, given in Eq.~\ref{eq:hz} will be finite. 
    The Jordan-Wigner transformation of $L^z_f$ will introduce a many body operator via the Jordan-Wigner string, $L^z_f=a^{\dagger}_{f}e^{i\pi\sum_{f}N_f}+h.c.$, and
the new fermion Hamiltonian will no longer be diagonalizable via a single particle transformation.}
%More generally and in general be non-integrable\spk{refs?}\cite{}.}
In this case, $n_q$ will no longer be conserved and, if $h^z$ is weak compared to the transverse field $h_f^x$, only the local conserved charges, $n_{f,0}^{c}$, $n_{f}^d$, and $N_{F}$, will survive after the first relaxation period.
If the disorder field, $h^{z}_{f}$, dominates over the transverse field, $h_f^x$, the effective Hamiltonian, Eq.~\ref{eq:effH2}, will many body localize and develop a set of local conserved charges $L^{z}_{f}$.
We have confirmed these expectations via exact diagonalization of the intermediate time Hamiltonian and by studying the level spacing statistics for $N_s=8$ and  $\delta N_s=24$ ($\delta=3$).

\section{Conclusions}
\label{sec:discussion}

A natural direction we are currently scrutinizing consists in extending the FE method to capture physics akin to the one reported in the experiment of Ref.~\cite{imma}. However, in order to have a quantitative understanding of   the delocalizing impact of the clean environment on the disordered chain, one should assume that the clean chain is delocalized, and therefore extend the ansatz employed here to treat Hamiltonian diagonal in momentum space.
It could  also be of interest to employ the FE method to study a broader variety of  MBL proximity effects. An appealing direction consists in studying a point-like, local coupling, between an MBL segment of interacting, disordered fermions and a clean one. This would pave way to understand the effect of the 'intrusion' of the localized system into the clean one, or viceversa, explore how an MBL system can act as an 'insulator' with respect to the clean segment.   Analysis in this direction is ongoing~\cite{Kellybis}.

\section*{Acknowledgments }

S. P. K. and J. M. are indebted with S. J. Thomson and M. Schiro for helpful and clarifying discussions and exchanges on the flow equation method for MBL systems.
We thank I. Bloch for inspiring discussions. 
JM is supported by the European Union's Horizon 2020 research and innovation
programme under the Marie Sklodowska-Curie grant agreement No 745608 (QUAKE4PRELIMAT). This research was supported in part by the National Science Foundation under Grant No. NSF PHY-1748958.

This work is based upon work supported in part (RN, JM) by the Air Force office of Scientific Research under award number FA 9550-17-1-0183.

S. P. K. acknowledges financial support from the UC Office of the President through the UC Laboratory Fees Research Program, Award Number LGF-17- 476883

Los Alamos National Laboratory is managed by Triad National Security,
LLC, for the National Nuclear Security Administration of the U.S.
Department of Energy under Contract No. 89233218CNA000001

\section*{References}
\bibliography{LRMBL}
\appendix

\section{Notation}
\label{apx:notation}
We define the onsite fields before Wick ordering as $h_{k}^{c(d)}$, and after Wick ordering, the effective fields are defined with a bar: $\bar{h}_{k}^{c(d)}$.
We define the couplings with unaccented variables with subscripts indexing sites: $\Delta^{I}_{ij}$, $\Delta^{c(d)}_{ij}$, $J_{ij}^{c(d)}$.
The dependence on the scale $l$ of the flow equations is often made implicit in expressions:$\Delta^{I}_{ij}(l)\rightarrow\Delta^{I}_{ij}$.
For $\Delta_{ij}^I$ the first index $i$ labels the clean chain sites and the second the dirty chain sites.
The spatial dependence of the couplings defines geometry and the magnitude is set by the parameters $\Delta^{I(c,d)}$, $J^{c(d)}$.
In addition to these parameters, the dirty chain fields are randomly selected from a box distribution, $[-W,W]$, and the Wick ordered reference state is set by: $\left<n^{d}\right>=\frac{1}{N_{s}}\sum_{k}\left<n^{d}_{k}\right>$ and temperature $\Theta$, where $\left<n_{k}^d\right>=\tr[\rho  n_{k}^{d}]$

We work with a set of unaccented operators: $U, H,H_0,V, H^c, H^d, H^I, \eta_{h}, \eta_{\Delta}, \eta_{I}, c_{k}, d_{k}, n^c_{k}, n_{k}^{d}$ and $\vec{L}_f=\{L^{x}_f,L^{y}_f,L^{z}_{f}\}$.
We also define a set of operators accented with a hat as:
\begin{eqnarray}
    \hat{J}^{c}=\sum_{ij}J^c_{ij}:c^{\dagger}_{i}c_{j}: \\ \nonumber
    \hat{J}^{d}=\sum_{ij}J^d_{ij}:d^{\dagger}_{i}d_{j}: \\ \nonumber
    \hat{\Delta}^{c}=\sum_{ij}\Delta^c_{ij}:n^c_{i}n^c_{j}: \\ \nonumber
    \hat{\Delta}^{d}=\sum_{ij}\Delta^d_{ij}:n^d_{i}n^d_{j}: \\ \nonumber
    \hat{\Delta}^{I}=\sum_{ij}\Delta^I_{ij}:n^c_{i}n^d_{j}: \\ \nonumber
    \hat{h}^{c}=\sum_{k}\bar{h}^c_{k}:n_{k}^{c}: \\ \nonumber
    \hat{h}^{d}=\sum_{k}\bar{h}^d_{k}:n_{k}^{c}:.
\end{eqnarray}

Finally, we also defined a symmetry operation, $C\leftrightarrow D$, that swaps the superscripts $c$ and $d$ of the couplings and operators and swaps the site indices of the inter-chain coupling:
\begin{eqnarray}
    &c \leftrightarrow d&\\ \nonumber
    &\Delta^{I}_{ij}\leftrightarrow {\Delta}^{I}_{ji}& \\ \nonumber
\end{eqnarray}

\section{Flow Equation Generators}
\label{apx:generators}
In the main text we defined 3 different generators the commutator: $\eta=[H_{0},J]=\eta_{h}+\eta_{\Delta}+\eta_{I}$ where:
\begin{eqnarray}
    \eta_{h}    &=& [\hat{J}^c,\hat{h}^c]        +  [\hat{J}^d,\hat{h}^d]       \\ \nonumber
    \eta_{\Delta} &=& [\hat{J}^c,\hat{\Delta}^{c}] +  [\hat{J}^d,\hat{\Delta}^{d}]\\ \nonumber
    \eta_{I}    &=& [\hat{J}^c+\hat{J}^{d},\hat{\Delta}^I]. 
\end{eqnarray}
and presented their form as:
\begin{eqnarray}
    \eta_{h}       &=& \sum_{ij}F^c_{ij}:c_{i}^{\dagger}c_{j}: +~ C\leftrightarrow D \\ \nonumber
    \eta_{\Delta}  &=& \sum_{ijk}\Gamma^c_{ij|k} :n^c_{k}c^{\dagger}_{i}c_{j}: + F^{\Delta^c}_{ij}:c_{i}^{\dagger}c_{j}:+~ C\leftrightarrow D \\ \nonumber
    \eta_{I}      &=& \sum_{ijk}\Gamma^{I}_{ij|k}:n^d_{k}c^{\dagger}_{i}c_{j}: + ~C\leftrightarrow D. \\ \nonumber
\end{eqnarray}
The coefficients $F$ and $\Gamma$ are given as:
\begin{eqnarray}
    F^c_{ij} &=& J^c_{ij}(\bar{h}^c_{i}-\bar{h}^c_{j}) \\ \nonumber
    F^{\Delta^c}_{ij} &=& 2J^c_{ij}\Delta^c_{ij}(n_{i}-n_{j}) \\ \nonumber
    \label{eq:Fij}
\end{eqnarray}
and
\begin{eqnarray}
    \Gamma^c_{ijk}&=&2J^c_{ij}(\Delta^c_{ik}-\Delta^c_{jk}) \\ \nonumber
    \Gamma^{I}_{ijk}&=&J^{c}_{ij}(\Delta^{I}_{ik}-\Delta^{I}_{jk}). \\ \nonumber
\end{eqnarray}
While the coefficient for the dirty chain can be obtained from the symmetry operation $C\leftrightarrow D$.

\section{The Flow Equations.}
\label{apx:floweq}
The full set of flow equations used in the numerics discussed in the main text is given as:
\begin{eqnarray}
    \frac{d\bar{h}_{k}^{c}}{dl}&=& \sum_{i}2(J^c_{ik})^2\left[(\bar{h}^c_{k}-\bar{h}^c_{i})+2\Delta^c_{ik}(n^c_{k}-n^c_{i})\right] \\ \nonumber
     && +2\sum_{ij}(J^c_{ij})^{2}(\Delta^c_{kj}-\Delta^c_{ki})(n^c_{j}-n^c_{i}) + \sum_{ij}(J_{ij}^{d})^{2}(\Delta_{kj}^{I}-\Delta_{ki}^{I})(n^{d}_{j}-n^{d}_{i}) \\ \nonumber
     \frac{dJ_{ij}^{c}}{dl} &=& \blue{-J^c_{ij}(\bar{h}^c_{i}-\bar{h}^c_{j})^{2}}-2J^c_{ij}\Delta^c_{ij}(n^c_{i}-n^c_{j})(\bar{h}^c_{i}-\bar{h}^c_{j})-\sum_{k}J^c_{ik}J^c_{kj}(2\bar{h}^c_{k}-\bar{h}^c_{i}-\bar{h}^c_{j}) \\ \nonumber
     &&-2\sum_{k}J^c_{ik}J^c_{kj}[\Delta^c_{ij}(n^c_{i}+n^c_{j}-2n^c_{k}) + 2\Delta^c_{ki}(n^c_{k}-n^c_{i}) + 2\Delta^c_{kj}(n^c_{k}-n^c_{j})] \\ \nonumber
     &&-J^c_{ij}(\bar{h}^c_{i}-\bar{h}^c_{j})(n^c_{i}-n^c_{j})(\Delta^c_{ij}+\Delta^c_{ji}) \\ \nonumber
 \frac{d\Delta_{ij}^{c}}{dl}&=&2\sum_{k\neq i,j l=i,j}(J^c_{lk})^{2}(\Delta^c_{ij}-\Delta^c_{kl'}) \\ \nonumber
     \frac{d\Delta_{ij}^{I}}{dl} &=&2\sum_{k}(J_{jk}^{d})^2(\Delta^{I}_{ij}-\Delta^{I}_{ik})+2\sum_{k}(J_{ik}^{c})^2(\Delta^{I}_{ij}-\Delta^{I}_{kj})
     \label{eq:flowEq}
\end{eqnarray}
where $n^{c(d)}_k=\left<n^{c(d)}_{k}\right>$ are the densities of the Wick ordered reference state, and the flow for the dirty couplings can be found using the symmetry operation $C\leftrightarrow D$.

\section{Numerical Details}
\label{apx:numerics}
The flow equations are numerically solved using an adaptive step $4^{th}$ order Runge-Kutta.
We work with a clean chain length of 24 sites $\delta N_s=24$ for a total of $48$ sites ($32$ sites when $\delta=3$).
We control the adaptive step by attempting around $800$ discrete Runge-Kutta steps on a log scale from $l=10^{-3}$ to $l=10^2$.
The adaptive step usually requires additional steps to reach the desired accuracy result in an average number of steps of around $3000$.

Since our results requires an accuracy for the couplings on a scale absolute scale $10^{-15}$, we devoted careful attention to numerical errors.
We found that numerical errors were due to floating-point errors for numbers close to $0$ during both the first step and at latter steps.
Numerical errors in the first step of a Runge-Kutta approximation are well-known, while the ones at later steps are due to the form of the flow equations.
These long time error are due to contributions like $\sum_{k}J_{ik}J_{kj}(h_{i}+h_{j}-h_{k})$ that could easily flip sign and cause numerical noise at longer times during the flow.

To manage these errors, we initialized the hoppings $J_{ij}$ for $i\neq j\pm1$ to $\epsilon_1$ and treated a hopping with $|J_{ij}|<\epsilon_2$ as exactly $0$.
Choosing $\epsilon_2>10^{-15}$ and $\epsilon_1>\epsilon_2$ was sufficient to reduce floating-point errors to the desired accuracy $10^{-15}$.
We tested the validity of these numerical approximations by varying $\epsilon_1$ and $\epsilon_2$ and observing no change in the flow.

\end{document}